\documentclass[final,5p,times,twocolumn]{elsarticle}
\usepackage[utf8]{inputenc}
\usepackage[english]{babel}
\usepackage{placeins}
\usepackage{xcolor}
\usepackage{geometry}                		
\usepackage{graphicx}				

\graphicspath{ {./Figures/}  }

\usepackage{graphicx}                 
\graphicspath{ {./Figures/} }

\usepackage{epstopdf}                 
\DeclareGraphicsExtensions{.png,.pdf,.jpg}  

\usepackage{booktabs}
\usepackage[backref=page,         
            pagebackref=false,    
            hyperindex=true,      
            colorlinks=true,      %
            citecolor=blue,       %
            filecolor=black,      %
            linkcolor=blue,       %
            urlcolor=blue,        %
            breaklinks=true,      %
            bookmarks=true,       
            bookmarksopen=false,  
            pdftitle   = {Second Measurements of Beam Backgrounds at the SuperKEKB Electron-Positron Collider},
            pdfauthor  = {The BEAST2 group},
            pdfsubject = {The subject}
]{hyperref}                       

\usepackage{amssymb}
\usepackage{textcomp}
\usepackage{siunitx}
\usepackage{multirow}
\def\SR {SR}

\begin{document}
\title{Measurements of Beam Backgrounds in SuperKEKB Phase 2}

\author[add10]{Z.~Liptak} 
 \author[add2,add3]{A.~Paladino}
\author[add8]{L.~Santelj}
\author[addUH]{J.~Schueler}
\author[addDesy]{S.~Stefkova}  
\author[add4]{H.~Tanigawa}      
\author[addMeidai]{N.~Tsuzuki}        
\author[addNapoli]{A.~Aloisio}
 \author[addBonn]{P.~Ahlburg}   
\author[addParis]{P.~Bambade}
\author[add13]{G.~Bassi\fnref{add14}}
 \author[addIPNS]{M.~Barrett}
 \author[add7]{J.~Baudot}
 \author[addUH]{T.~E.~Browder}
 \author[add2,add3]{G.~Casarosa}
 \author[add13,add15]{G.~Cautero}
 \author[add11]{D.~Cinabro}
 \author[add7]{G.~Claus}
 \author[add7]{D.~Cuesta}
\author[addNapoli]{F.~Di~Capua}
\author[addParis]{S.~Di~Carlo}
 \author[addSKB]{J.~Flanagan}
 \author[add5]{A.~Frey}
 \author[add9]{B.~G.~Fulsom}
 \author[addSKB]{Y.~Funakoshi}
 \author[addMPPI]{M.~Gabriel}
\author[addNapoli]{R.~Giordano}
 \author[add13,add15]{D.~Giuressi}
 \author[add7]{M.~Goffe}
 \author[addKEK,add12]{K.~Hara}
 \author[addUH]{O.~Hartbrich}
 \author[addUH]{M.~T.~Hedges\corref{cor1}\fnref{purdue}}
\fntext[purdue]{Now at Purdue University, West Lafayette, IN 47907, USA}
 \author[addMPPI,add19]{D.~Heuchel}
  \author[addSKB]{N.~Iida}
 \author[addSKB]{T.~Ishibashi}
 \author[add7]{K.~Jaaskelainen}
  \author[addParis]{D.~Jehanno}
\author[addUVic]{S.~de~Jong}
\author[addMPPI]{T.~Kraetzschmar}
\author[add13]{C.~La~Licata\fnref{add16}}
\author[add13]{L.~Lanceri}
\author[addMPPI]{P.~Leitl}
\author[addUH]{P.~M.~Lewis\fnref{bonn}}
\fntext[bonn]{Now at Universit\"at Bonn, 53115 Bonn, Germany}

\author[add6]{C.~Marinas}
\author[addUVic]{C.~Miller}
 \author[addMPPI]{H.~Moser}
 \author[addKEK,add12]{K.~R.~Nakamura}
 \author[addKEK]{H.~Nakayama}
 \author[addDesy]{C.~Niebuhr}
  \author[add4]{Y.~Onuki}
\author[addParis]{C.~Pang}
 \author[addBonn]{B.~Paschen}
 \author[add7]{I.~Ripp-Baudot}
 \author[add2,add3]{G.~Rizzo}
 \author[addUVic]{J.~M.~Roney}

 \author[add5]{H.~Schreeck}
 \author[add5]{B.~Schwenker}
 
\author[addMPPI]{F.~Simon}
 \author[add7]{M.~Specht}
 \author[addGut]{B.~Spruck}
 \author[addDesy]{Y.~Soloviev}

 \author[add7]{M.~Szelezniak}
 \author[addKEK,add12]{S.~Tanaka}
 
 \author[addSKB]{S.~Terui}
\author[addNapoli]{G.~Tortone}
 \author[addKEK,add12]{T.~Tsuboyama}

 \author[add4]{Y.~Uematsu}
 \author[addUH]{S.~E.~Vahsen}
 \author[add13]{L.~Vitale}
 \author[addMPPI]{H.~Windel}





\address[addBonn]{Universit\"at Bonn, 53115 Bonn, Germany}
\address[addDesy]{Deutsches Elektronen--Synchrotron, 22607 Hamburg, Germany}
\address[add15]{Elettra-Sincrotrone Trieste S.C.p.A., Strada Statale SS14, km 163.5, Basovizza 34149, Italy}
\address[add5]{Georg-August-Universit\"at G\"ottingen, II. Physikalisches Institut, 37073 G\"ottingen, Germany}
\address[add12]{The Graduate University for Advanced Studies (SOKENDAI), Hayama 240-0193, Japan}
\address[addKEK]{High Energy Accelerator Research Organization (KEK), Tsukuba, 305-0801, Japan}
\address[addSKB]{High Energy Accelerator Research Organization (KEK), Accelerator Laboratory, Oho 1-1, Tsukuba, Ibaraki, 305-0801, Japan}
\address[add10]{Hiroshima University, Higashi-Hiroshima, Hiroshima 739-8530, Japan}
\address[add2]{INFN Sezione di Pisa, I-56127 Pisa, Italy}
\address[addIPNS]{Institute of Particle and Nuclear Studies, High Energy Accelerator Research Organization (KEK), 1-1 Oho, Tsukuba, Ibaraki, 305-0801, Japan}
\address[add8]{J.~Stefan Institute, 1000 Ljubljana, Slovenia}
\address[addGut]{Johannes Gutenberg--Universit\"{a}t Mainz, Institut f\"{u}r Kernphysik,  55128 Mainz, Germany}
\address[addMeidai]{Graduate School of Science, Nagoya University, Nagoya 464-8602, Japan}
\address[addNapoli]{Univ. of Naples Federico II \& INFN Sezione di Napoli, Strada Comunale Cintia, 80126 Napoli, Italy}
\address[add9]{Pacific Northwest National Laboratory, Richland, Washington 99352, U.S.A.}
\address[add3]{Universit\`{a} di Pisa, Dipartimento di Fisica, I-56127 Pisa, Italy}
\address[add7]{Universit\'e de Strasbourg, CNRS, IPHC UMR 7178, Strasbourg, France}
\address[add13]{Universit\`{a} di Trieste Dipartimento di Fisica, and INFN Sezione di Trieste, I-34127 Trieste, Italy}
\address[add4]{University of Tokyo, Department of Physics, Tokyo 113-0033, Japan}
\address[add6]{University of Valencia - CSIC, Instituto de Fisica Corpuscular (IFIC), Spain}
\address[addUVic]{University of Victoria, Department of Physics and Astronomy, 3800 Finnerty Rd., Victoria BC, V8P 5C2, Canada}
\address[add11]{Wayne State University, Detroit, Michigan 48202, U.S.A.}
\address[addMPPI]{Max-Planck-Institut f\"ur Physik, F\"{o}ringer Ring 6, 80805 M\"unchen, Germany}
\address[addUH]{University of Hawaii, Department of Physics and Astronomy, 2505 Correa Road, Honolulu, HI 96822, U.S.A.}
\address[addParis]{Université Paris-Saclay, CNRS/IN2P3, IJCLab, 91405 Orsay, France}

\fntext[add14]{Now at Dipartimento di Fisica, Universit\`{a} di Pisa and INFN Sezione di Pisa, I-56127 Trieste, Italy}

\fntext[add16]{Now at Kavli Institute for the Physics and Mathematics of the Universe (WPI), University of Tokyo, Kashiwa 277-8583, Japan} 
\fntext[add19]{Now at DESY, 22607 Hamburg, Germany}

\date{}

\begin{abstract}
The high design luminosity of the SuperKEKB electron-positron collider will result in challenging levels of beam-induced backgrounds in the interaction region. Understanding and mitigating these backgrounds is critical to the success of the Belle~II experiment. We report on the first background measurements performed after roll-in of the Belle II detector, a period known as SuperKEKB Phase 2, utilizing both the BEAST II system of dedicated background detectors and the Belle II detector itself. We also report on first revisions to the background simulation made in response to our findings. Backgrounds measured include contributions from synchrotron radiation, beam-gas, Touschek, and injection backgrounds. At the end of Phase 2, single-beam backgrounds originating from the 4 GeV positron Low Energy Ring (LER) agree reasonably well with simulation, while backgrounds from the 7 GeV electron High Energy Ring (HER) are approximately one order of magnitude higher than simulation. We extrapolate these backgrounds forward and conclude it is safe to install the Belle II vertex detector.
\end{abstract}

\maketitle



\section{Introduction}
\index{Introduction}
\label{section_introduction}

The SuperKEKB asymmetric electron-positron collider \cite{PTEP:Ohnish} is currently in operation at KEK in Tsukuba, Japan, producing collision events for the Belle~II experiment \cite{Abe:2010gxa}. Its current target is a luminosity of \SI{6.5e35}{cm^{-2}s^{-1}}, a roughly 30-fold improvement over its predecessor, KEKB \cite{KEK:1995sta}. The increased luminosity is to be achieved by moving to a new ``nano-beam" scheme~\cite{Bona:2007qt}, which will reduce beams to a vertical size of \SI{50}{nm} at the interaction point (IP), one twentieth the size of its predecessor, while increasing the beam currents by approximately 50\%.

The reduced beam size, increased beam currents, and consequent increased luminosity will provide a challenging environment for physics measurements, as each improvement also increases the background at the IP. Thus, measuring and mitigating these backgrounds is of critical importance for successful physics data taking, particularly in the later stages of the experiment as SuperKEKB approaches its target luminosity.

In Section~\ref{sec:experimental_setup} we describe the detectors used for dedicated measurements of backgrounds, particularly those which were added or modified between the first and second SuperKEKB accelerator commissioning phases. In Section~\ref{sec:operational_experience} we describe the operational experience during Phase~2. Sections~\ref{sec:simulation} and~\ref{sec:exp_results} describe the simulation of relevant backgrounds and their measurements, respectively. Section~\ref{sec:collimator_study} describes the procedure for optimizing the collimator settings for the best compromise between beam lifetime and background levels. Finally, Section~\ref{sec:summary_recommendations} summarizes the results of Phase~2, extrapolations to expected levels of background for future Belle~II data taking runs, and recommendations for future operation.

\subsection{Operational Phases}
\index{Phases}
\label{sec:phases}
We divide Belle~II/SuperKEKB running into 3 major phases: Phase~1, carried out in the Spring of 2016, was the first of two dedicated commissioning phases. This phase was run without the Belle~II detector installed around the IP and also without final focusing of the accelerator and thus, no collisions. In place of Belle~II, a dedicated suite of background detectors collectively known as BEAST II was installed around the IP to directly measure rates of backgrounds produced in the accelerator. The results from Phase~1 are reported in Reference~\cite{Lewis:2018ayu}.

In Phase~2, most of the Belle~II detector assembly was installed around the IP, with the notable exception of the VerteX Detector (VXD), of which only one octant was installed. Phase~2 began in March 2018, with first electron-positron collisions in April 2018, and concluded in July 2018.
From the Belle~II point of view, the major goal of this phase was to confirm that the sensitive VXD detectors could be safely installed; for SuperKEKB, the goal was machine performance studies and achieving a target luminosity of $1 \times 10^{34} \mathrm{cm}^{-2}\mathrm{s}^{-1}$.

Phase~3 refers to the main physics data-taking run of the Belle~II experiment, encompassing the remainder of the lifetime of the project. This includes the full operation of all Belle~II subdetectors. During this phase, SuperKEKB plans to gradually improve perfomance up to its design luminosity, eventually accumulating 50 $\mathrm{ab}^{-1}$ of collision data.

\subsection{Beam-Related Backgrounds and Their Origins }
\label{sec:types}

Here, we provide an overview of the major sources of backgrounds incident onto Belle~II from the SuperKEKB beamline and their expected dependence on accelerator conditions.  Using these expected dependencies, we then build a background model that we use later to extract the relative contributions to the total background rates in BEAST II and Belle~II subdetectors. 

We refer to backgrounds arising from Coulomb interactions between particles in the same beam bunch as ``Touschek backgrounds''. Interacting beam particles either gain or lose energy and thus, subsequent to scattering, propagate with an energy higher or lower than the nominal bunch energy. Affected particles will deviate from the nominal orbit and eventually collide with the beam pipe wall. Showers from such collisions near the IP may reach a subdetector and produce unwanted background events. We expect the rate of Touschek scattering to be inversely proportional to the number of filled bunches in the main storage ring (MR), $n_b$, and the square of the beam current $I$, and inversely proportional to the bunch volume $\sigma_{x} \times \sigma_{y} \times \sigma_{z}$, where $x$ and $y$ refer to horizontal (in the plane of the rings) and vertical directions, respectively, and $z$ is the direction of propagation of the beam~\cite{PTEP:Ohnish}. 

During operation, particles in the beams occasionally interact with residual free gas molecules inside the beam pipe. As described for Touschek scattering, resulting off-orbit beam particles can hit the beam pipe wall near the IP and cause background events, which we refer to as beam-gas events. Because the likelihood of interactions with gas nuclei increases as more gas molecules are present, the beam-gas event rate scales linearly with the beam pipe pressure.  

In addition to a baseline level of gas present in the beam pipe, the pressure also tends to rise during operation as the beam causes increased outgassing from the beam pipe material. In the long term, the pressure tends to decrease as the total integral of the beam current increases; hence, the newer LER beamline exhibits significantly larger pressure, and therefore more beam-gas events, than the HER.

As electrons and positrons are accelerated around the circular MR, emission of synchrotron radiation (SR) is unavoidable. SR photons incident on Belle~II are typically in the range of a few to several tens of\,keV. To protect the innermost detectors against these photons, the inner surface of the central 2\,cm-diameter Beryllium beam pipe will be coated with a 10\,\textmu m-thick layer of gold. In order to enhance the sensitivity to these SR photons during Phase~2 the thickness of the gold layer was reduced to 6.6\,\textmu m. Dedicated studies for understanding SR backgrounds are described in Section \ref{sec:synchrotron_sim}. Because SR power is proportional to the square of both beam energy and magnetic field strength, we expect the HER to be the dominant source of SR production.

Due to the short lifetime of the circulating beams, we must inject beam even when beams are in collision and Belle~II is taking data. When beam is injected from the SuperKEKB linac into the MR, the injected bunch is of lower quality than that of the stored beam and causes a perturbation that can last several milliseconds and is observable as an increased background rate for many turns around the ring. Reducing the emittance of the injected beam is essential both to mitigate this background as well as for reaching high luminosity, so a positron damping ring was installed prior to Phase~2 to reduce emittance in the LER and provide a cleaner beam. We also use a dedicated injection background detector, sCintillation Light and Waveform Sensors (CLAWS), specifically to monitor injection background waveforms and assist in determining the necessary post-injection veto window to avoid excessive occupancy in the  PiXel Detector (PXD). We describe a dedicated injection background study in Section \ref{sec:injection_backgrounds}.

We also consider backgrounds that arise from beam-beam collisions. Unlike the effects already described, these backgrounds require the beams to interact at the IP, and are hence referred to as luminosity backgrounds. Because the beam conditions during Phase~2 involved low currents and large beam sizes relative to the final design parameters, single beams were the primary sources of beam-related backgrounds. As currents are increased and the beam approaches the full nano-beam scheme, we expect luminosity backgrounds to overtake single-beam backgrounds.


\section{Experimental Setup}
\label{sec:experimental_setup}
\index{Experimental Setup}

The experimental setup for Phase~2 can be divided into three distinct systems: the SuperKEKB accelerator facility; Belle~II; and BEAST II, the collective name for the dedicated beam-background detectors.

\subsection{SuperKEKB}

\begin{figure}
	\begin{center}
		\includegraphics[width=0.5\textwidth]{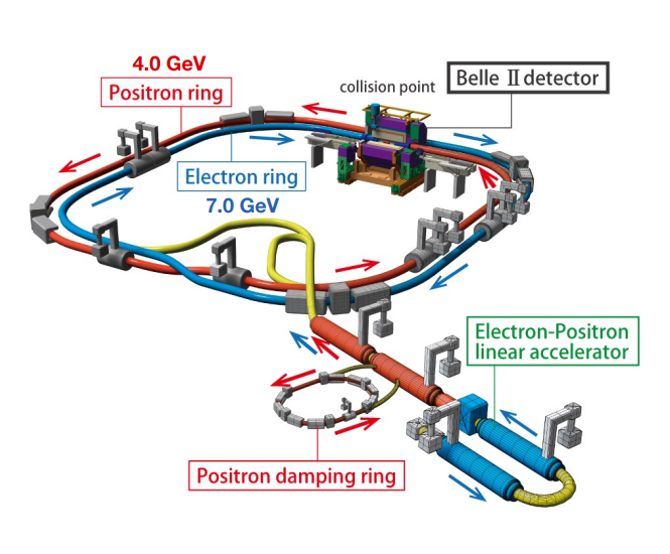}
		\caption{Overview of the SuperKEKB accelerator facility, showing the Belle~II detector installed at the interaction point.}
                  
		\label{fig:skb_layout}
	\end{center}
\end{figure}

SuperKEKB is the asymmetric electron-positron collider situated at the High Energy Accelerator Research Organization (KEK) in Tsukuba, Japan. It consists of an approximately 3\,km MR fed by a linear accelerator via beam transport lines, as shown in Figure~\ref{fig:skb_layout}. The MR is composed of a low energy ring providing 4\,GeV positrons and a high energy ring that provides 7\,GeV electrons. Further details concerning the physical setup of SuperKEKB can be found in Reference~\cite{PTEP:Ohnish}. 

\subsection{Accelerator Parameter Monitors} 

Understanding the accelerator and beam conditions is of critical importance in the simulation and measurement of beam-related backgrounds. In particular, our model of backgrounds at the IP relies on the following beam parameters, as described in Section~\ref{sec:heuristic}: vertical beam size, $\sigma_{y}$, gas pressure $P$, gas composition $Z_{\text{eff}}$, and beam current $I$. This section describes in brief the apparatus used to measure the quantities listed above. More detailed descriptions may be found in Reference~\cite{Lewis:2018ayu}.

\subsubsection{Beam Size}

Horizontal and vertical measurements of the beam sizes in both the LER and HER are performed with X-ray monitors (XRM) as well as visible light monitors, implemented in the LER and HER late in Phase~2. Beam size measurements reported here are taken from the XRM system, described in References~\cite{Mulyani:IBIC2015-TUPB025} and~\cite{IBIC2016}.

\subsubsection{Beam Pipe Gas Pressure}
\label{sec:Beam_Gas_Pressure}

As was the case in Phase~1, beam pipe gas pressure is a large uncertainty in the analysis.  Gas pressure is measured with cold cathode gauges distributed approximately every 10\,m around each ring. Beam pipe gas pressure can vary substantially throughout the ring and changes far upstream of the IP can greatly influence the background rates seen in Belle~II detectors.

The beam-gas pressure is composed of a base value present in the pipe at all times, typically less than 10$^{-9}$\,Pa, and a dynamic component that appears only while the beam is in operation. The dynamic pressure component was found to increase approximately linearly with beam current. 

\subsubsection{Beam-Gas Composition}

To improve the modeling of beam-gas interactions, data from three Residual Gas Analyzers (RGAs) is used to build a comprehensive model of gas species in the beam pipe. The RGAs provide partial pressures of individual mass species, which are then combined to give an effective $Z$ value, $Z_{eff}$, for the gas mixture in the pipe. In addition to the two RGAs used in Phase~1, a third RGA was added to the HER for Phase~2. 

\subsection{Belle~II}

Beginning in Phase~2 the full set of Belle~II subdetectors, with the exception of most of the sensitive VXD consisting of a PXD and Silicon Vertex Detector (SVD), was installed around the IP. Surrounding the single installed octant of the VXD are a Central Drift Chamber (CDC) for particle tracking, a Time-Of-Propagation (TOP) counter and Aerogel Ring Imaging Cherenkov Device (ARICH) for particle identification, and a CsI(Tl)-based Electromagnetic CaLorimeter (ECL). This assembly is enclosed in a solenoidal magnet for generation of a 1.5\,T magnetic field. An outer detector for long-lived $K_{L}$ and $\mu$ particles consisting of a barrel with two layers of scintillators surrounded by a resistive plate counter and two endcaps composed entirely of scintillators is situated farthest outward radially. 

\begin{figure}[htb]
	\begin{center}
		\includegraphics[width=.45\textwidth]{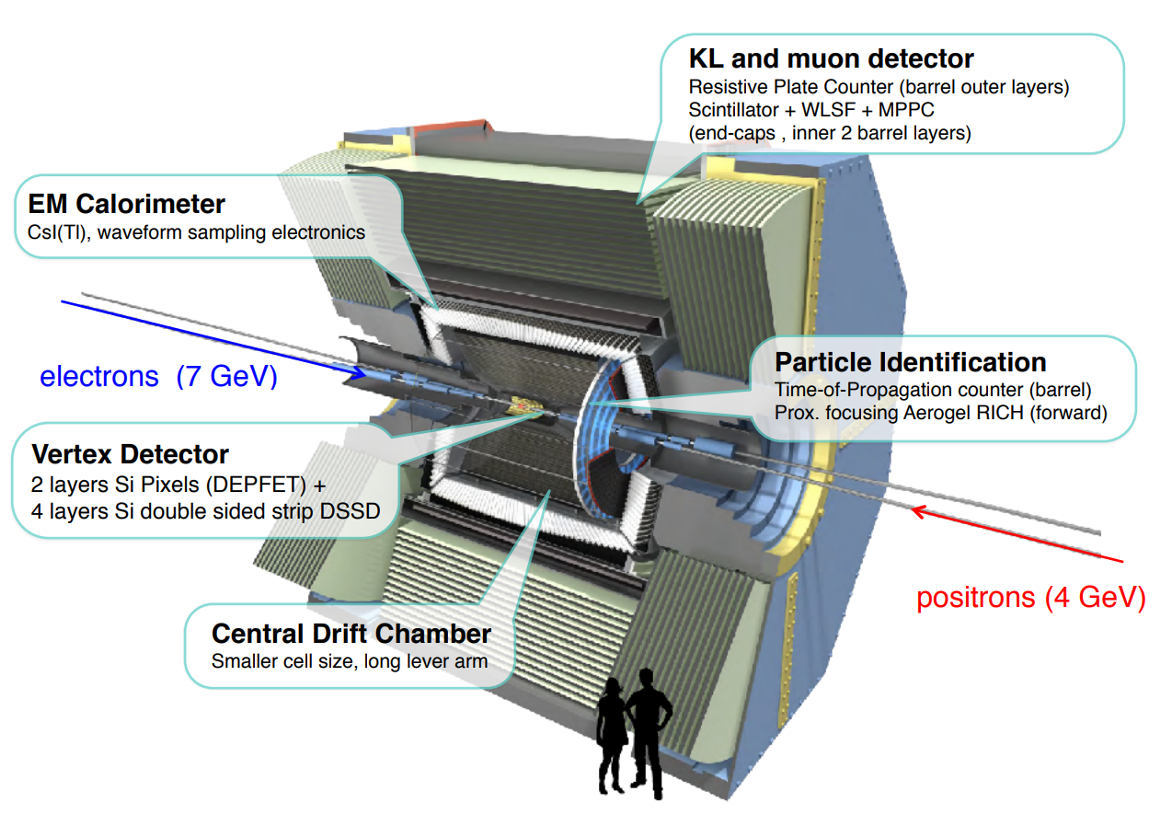}
		\caption{Cutaway diagram of the Belle~II detector, with individual subdetector systems labeled in word balloons. The VerteX Detector (VXD) system was only partially installed, with only one octant present during SuperKEKB Phase 2.}
		\label{fig:belle_ii_detector}
	\end{center}
\end{figure}

\begin{figure}[htb] 
    \begin{center}
      
      \includegraphics[width=.45\textwidth]{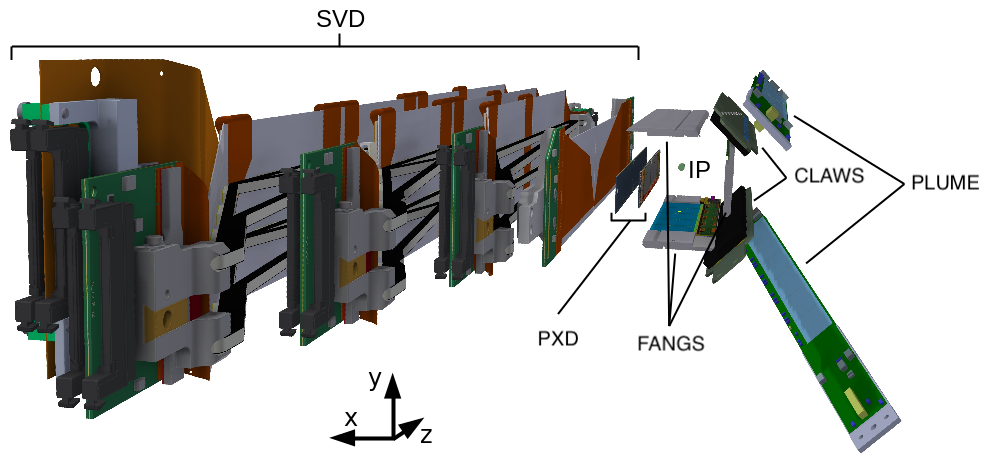}
      \caption{Vertex detector configuration during Phase 2. Only one octant of the nominal Belle~II vertex detector, composed of the SVD and PXD, was installed. This allowed nominal Belle~II reconstruction of select, horizontally-going charged tracks from the IP. Particles going in other directions were monitored with dedicated BEAST II background monitoring detector systems (FANGS, CLAWS, and PLUME).}
      
        \label{fig:beast:vxd_arrangement}
    \end{center}
\end{figure}

The full Belle~II experimental setup, including the VXD, which was only partially installed for Phase~2, is shown in Figure~\ref{fig:belle_ii_detector}. More information about the Belle~II experimental setup can be found in Reference~\cite{Abe:2010gxa}. The inner detector configuration for Phase~2 is shown in detail in Figure~\ref{fig:beast:vxd_arrangement}.

\subsection{BEAST II Detectors}

During Phase~1, we also deployed a suite of background detectors collectively known as BEAST II~\cite{Lewis:2018ayu}. Each detector was designed to measure a particular type of background expected during Belle~II physics operations, including X-rays, charged particles, and fast and thermal neutrons.

In the transition from Phase~1 to Phase~2, the BEAST II support structure at the IP was removed, and replaced by the rolled-in Belle~II detector. Many of the original BEAST II detectors were re-used and moved to new locations in Phase~2, to continue the monitoring of specific background components. For example, CLAWS injection background monitors were moved very close to the IP, and BEAST fast neutron detectors were installed around the final focusing Superconducting Quadrupole (QCS) magnets on either side of the IR, where beam losses and resulting backgrounds are particularly high. Because a primary goal of Phase~2 was to determine if it was safe to install the full Belle~II VXD, only one $\phi$ (azimuthal angle) segment replica of that detector was installed. The remainder of the VXD volume was instrumented with three types of custom background detectors; FANGS, PLUME, and the new CLAWS configuration, all described below. The complete set of custom background detectors, both those remaining from Phase~1, and those new in Phase~2, will be collectively referred to as ``BEAST II detectors''. The reason for this designation is that these detectors are not integrated into the Belle~II DAQ stream, but rather monitored separately.

\subsubsection{FANGS Detector System} 
The FE-I4 ATLAS Near Gamma Sensors (FANGS) detector was specifically designed for BEAST II. It is based on hybrid pixel detector modules used in the ATLAS pixel detector (FE-I4), which are sensitive to charged tracks and low-energy ($\approx$\,10\,keV) X-rays. They are designed to withstand radiation of up to 300\,Mrad and to cope with a maximum hit rate of 400\,MHz/cm$^2$~\cite{FE-I4:2012}.\par
\begin{figure}[phtb]
    \begin{center}
        \includegraphics[width=\linewidth]{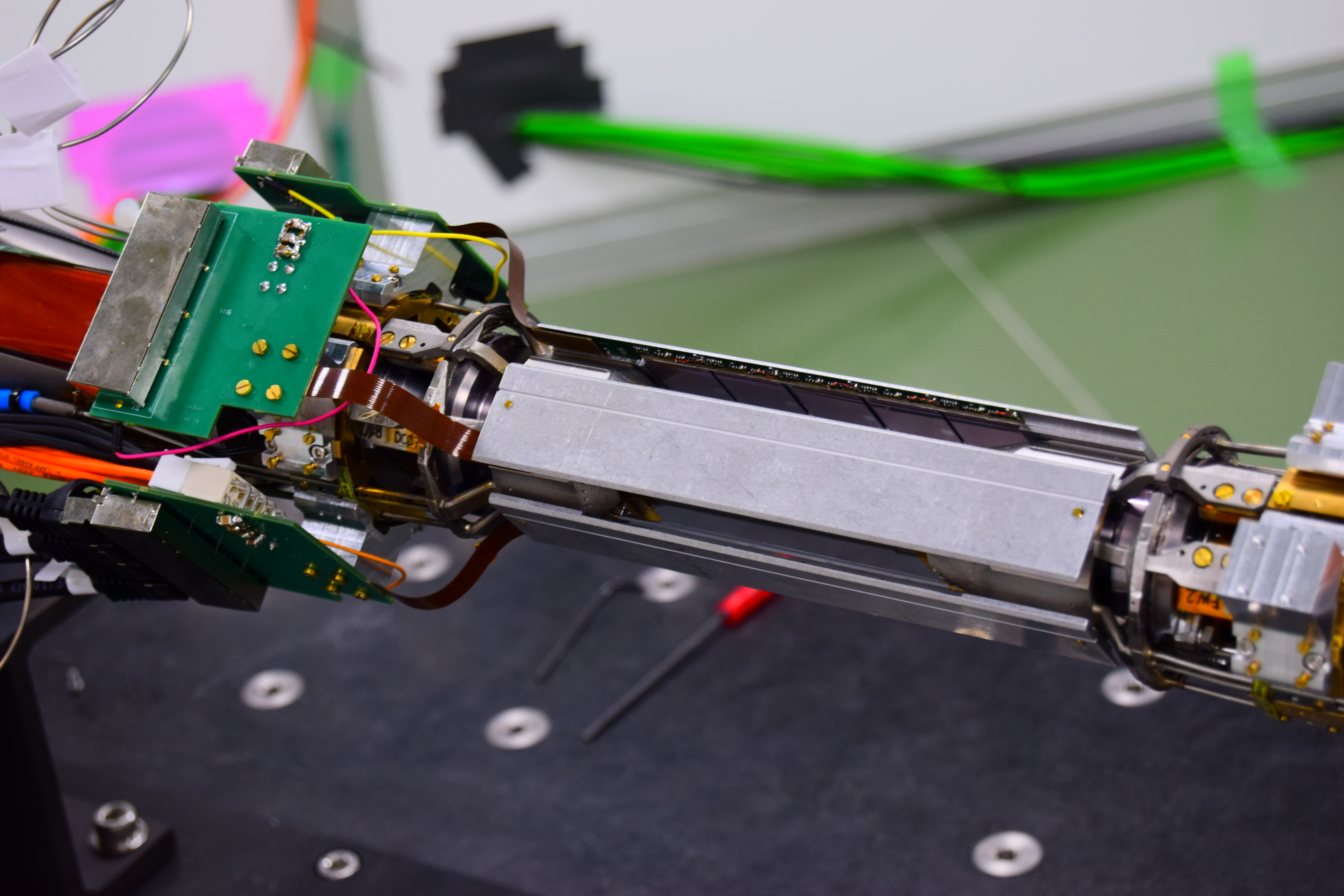}
        \caption{The three FE-I4 ATLAS Near Gamma Sensors (FANGS) staves, utilizing hybrid pixel sensors, installed around the beam pipe. This photo was taken before installation of other vertex detectors.}
     
        \label{fig:fangs:staves}
    \end{center}
\end{figure}

The basic unit of the FANGS detector system is the stave, of which three are installed around the beam pipe (see \autoref{fig:fangs:staves}). Each stave contains five FE-I4 readout chips connected to 250\,\textmu m thick n-in-n planar silicon sensors, mounted on an aluminum profile. Each stave is mounted on two cooling blocks at a distance of 22\,mm from the interaction region, at azimuthal positions of $90^{\circ}$, $180^{\circ}$, and $270^{\circ}$ in $\phi$. The distance between the detector and the back-end electronics is 18\,m.\par

In BEAST II Phase~2, the system provides hit rates as well as the recorded charge spectrum from ionizing radiation. More details about the FANGS detector system can be found in Reference~\cite{Ahlburg:2016}.

\subsubsection{CLAWS Injection Background Detector System} 
The CLAWS system targets the beam background arising from injections of new particles into the main ring. Its sub-nanosecond time resolution allows measurement of the exact arrival time of minimum-ionizing particles (MIPs) emerging from bunches crossing the IP. Recording single waveforms over a period of multiple tens of milliseconds gives important insights about the time structure of injected bunches and overall beam behavior.
\par
\begin{figure}[phtb]
    \begin{center}
              \includegraphics[width=\linewidth]{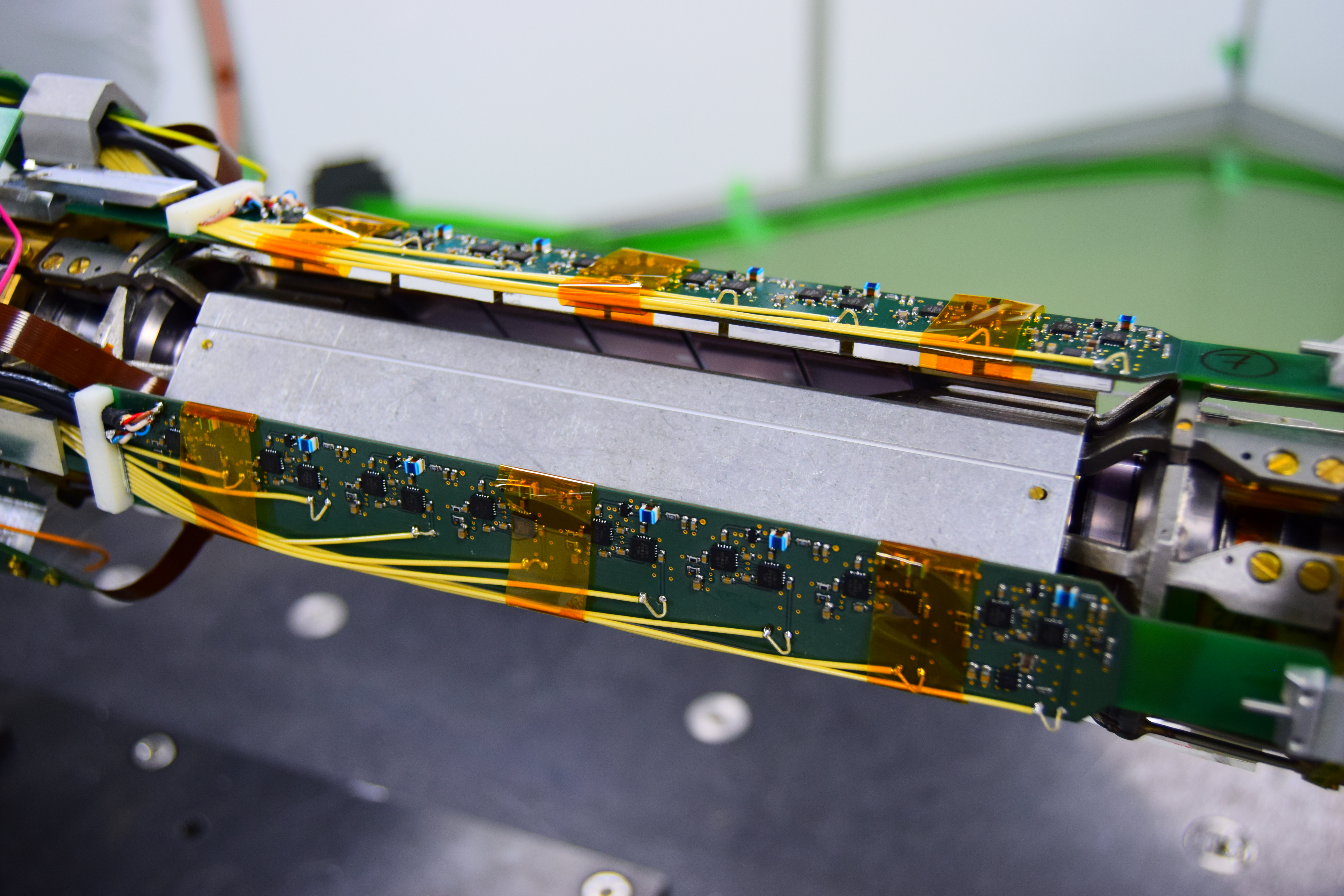}
        \caption{Two CLAWS staves, based on plastic scintillator tiles instrumented with silicon photomultipliers, installed at $\phi = 135^{\circ}$ and 225$^{\circ}$, which correspond to the top and bottom of the figure, respectively. One FANGS stave is also visible between the two CLAWS staves at $\phi = 180^{\circ}$.}
          
        \label{fig:claws:ladder_example}
    \end{center}
\end{figure}

The active part of the detector consists of plastic scintillator tiles instrumented with silicon photomultipliers (SiPMs), comparable to the CLAWS sensors used in Phase~1~\cite{Lewis:2018ayu}. The scintillator tiles are arranged in two staves with eight independent tiles each, as shown in Figure~\ref{fig:claws:ladder_example}. The sensors, which are mostly sensitive to charged particles, replace two modules of the second layer of the PXD as depicted in Figure~\ref{fig:beast:vxd_arrangement}. The switchable on-board pre-amplifiers enable a high and low gain mode. In addition, a second amplifier is located in a dockbox just outside of the VXD volume, increasing the amplitude of the analog signal before its transmission over a distance of $35\,\mathrm{m}$ to the DAQ.

\subsubsection{PLUME detector system}

The Pixelated Ladder with Ultra-low Material Embedding (PLUME) detector is designed as a hit rate monitor, with a material budget low enough to minimally impact particle trajectories~\cite{Cuesta:2020bpl}. It is based on ultra-light double-sided ladders equipped with CMOS pixel sensors, initially developed as a generic concept for the vertex detector of the International Large Detector (ILD)~\cite{Nomerotski:2011zz}. Taking advantage of its very low material budget of 0.4\% of $X_0$, two PLUME ladders are installed in the VXD volume. Both units feature a sensitive area of $127\times11$~mm$^2$ (for each side), with their long axis oriented along the beam. One ladder sits parallel to the beam axis, at $\phi=125^{\circ}$, and the second one is tilted at an angle of $\theta=18^{\circ}$ with respect to the beam, at $\phi=225^{\circ}$. This arrangement, depicted in Figure~\ref{fig:beast:vxd_arrangement}, was chosen in order to monitor the beam-induced background in a range of radii from 5 to 9\,cm with only two devices.

\begin{figure}
\includegraphics[width=.45\textwidth]{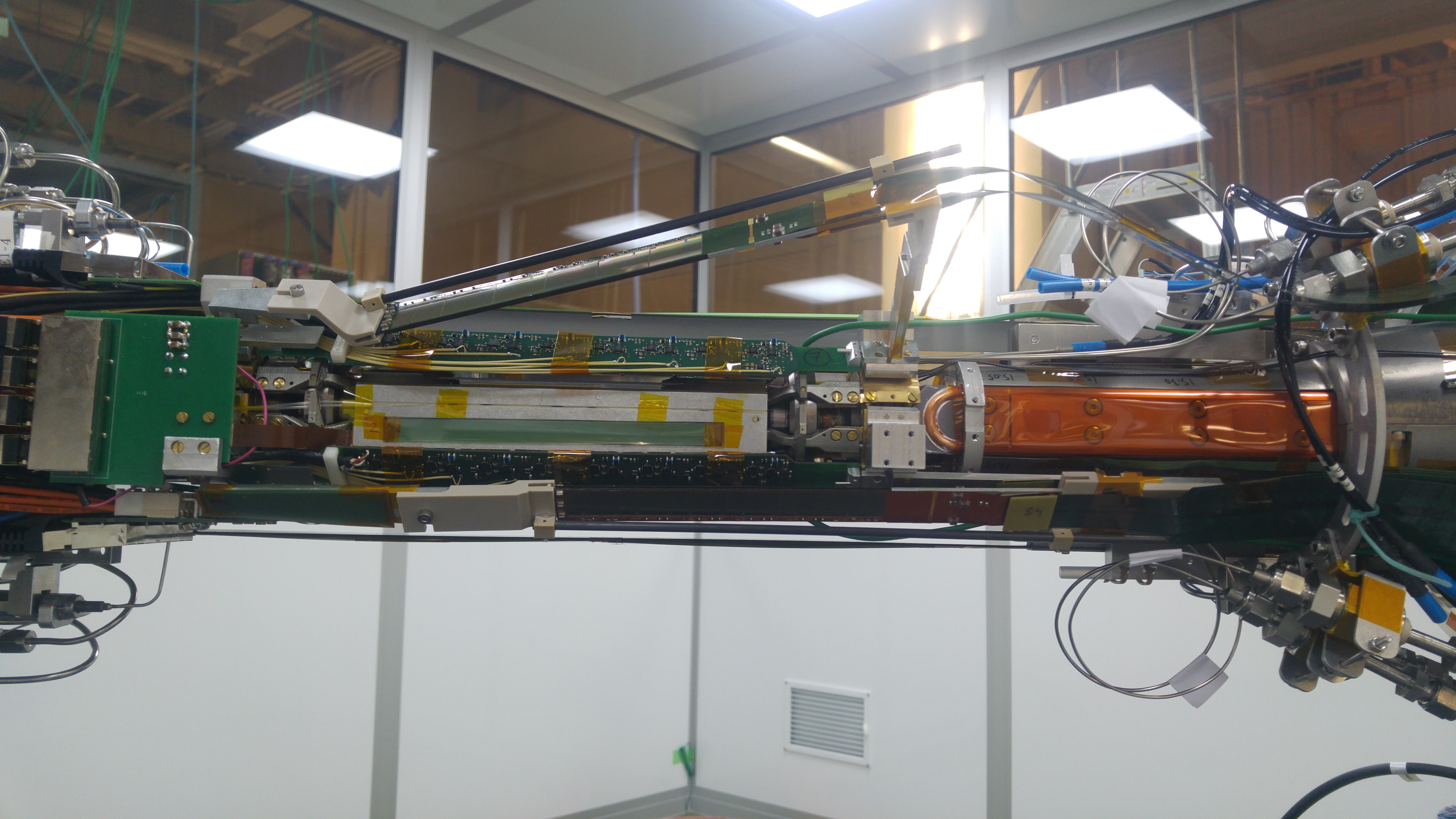}
\caption{BEAST II vertex detector assembly after installation of the two CMOS pixel sensor-based PLUME staves. The PLUME staves are visible at the top and left hand side of the assembly.}
  \label{fig:PLUME}

\end{figure}

Each ladder includes 6 MIMOSA-26 CMOS pixel sensors~\cite{Baudot:2013pca} on each side, with a total of 8$\times$10$^6$ pixels per ladder. The sensor granularity and the frame-based read-out architecture, featuring a 115~\si{\micro}s integration time, allow a counting rate in excess of $10^6$\,hits\,cm$^{-2} \cdot $s$^{-1}$. Sensors are operated with a low detection threshold, which enables sensitivity to electrons with energies greater than about 40~keV and to X-rays in the 2-10~keV energy range due to the thin sensitive layer (14~\si{\micro}m). The data acquistion (DAQ) system is designed to read out both PLUME ladders with no dead time. The system provides pixel counts for every sensor frame separated by 115\,\si{\micro}s for use in studying the background evolution of newly injected bunches, and averages the counts over 1 second for standard online monitoring. The PLUME system is shown in Figure~\ref{fig:PLUME}.

\subsubsection{BEAST TPC Fast Neutron Detector System} 

\begin{figure}[phtb]
    \begin{center}
        \includegraphics[width=0.5\linewidth]{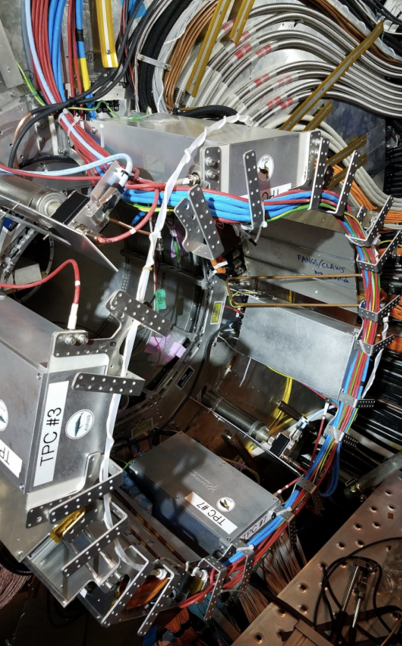}
        \caption{Photograph of the VXD dock space at $z = -\SI{1.3}{m}$, an area reserved for cabling and patch panels inside the Belle~II detector, immediately outside the CDC endplate. Four BEAST TPC fast neutron detectors are installed at different azimuthal angles ($\phi$=18$^\circ$, 90$^\circ$, 198$^\circ$, and 270$^\circ$). Two $^{3}$He thermal neutron detectors are are also visible at $\phi$ = 40$^\circ$ and 230$^\circ$. (Coordinates listed are in the global accelerator system; in this figure the positive $x$ axis is to the left and $\theta$ increases clockwise.)} 
         \label{fig:tpcs:geometry}
    \end{center}
\end{figure}

We use a system of eight identical, compact time projection chambers (TPCs)  with pixel ASIC readout to detect fast neutrons via their nuclear recoils in gas. The detectors were specifically designed for SuperKEKB commissioning~\cite{Lewis:2018ayu, Jaegle:2019jpx, Thorpe:2021qce} and provide the rate, energy spectra, and three-dimensional directions of fast neutron recoils inside the Belle~II detector. The TPCs are situated in the VXD dock spaces on either side of the IR, with four TPCs located at $z = -\SI{1.3}{m}$, as shown in Figure~\ref{fig:tpcs:geometry}, and the remaining four at $z = +\SI{1.9}{m}$.  These locations are critical for neutron monitoring, as during operation the VXD docks surround a part of the beam focusing system where the radius of the incoming beam pipe decreases. This results in particularly high beam particle loss rates, leading to high rates of background showers and neutrons. These neutrons can travel radially outward and penetrate the outer Belle~II detectors from the inside. Verifying that this background component agrees with expectation is therefore important, so that extra neutron shielding can be added if required. The detectors operate using a 70:30 He:CO$_2$ gas mixture.

Detailed descriptions of gas delivery, high voltage operation, and the data acquisition system in Phase~2 can be found elsewhere~\cite{Schueler:2021}. The added detector coverage in $\phi$ allows for a measurement of spatial distributions of fast neutron recoils that compliments the fast neutron-recoil angular distribution measurements previously performed in Phase~1~\cite{Hedges:2021dgz}.

\subsubsection{He-3 Thermal Neutron Detector System} 
During Phase~2 the $^3$He system monitored the thermal neutron flux and was capable of triggering an alarm upon detecting dangerous neutron levels. In preparation for Phase~2, the DAQ was upgraded to include a NIM discriminator module to improve counting efficiency, and extensive calibrations were carried out with a thermal neutron source. The detectors themselves are unchanged from Phase~1~\cite{Lewis:2018ayu}; however, they were moved to new positions to account for the presence of the Belle~II detector. Two detectors are mounted in the $+z$ dock space at 40$^{\circ}$ and 230$^{\circ}$ in $\phi$, as seen in Figure~\ref{fig:tpcs:geometry}, with a similar setup used in the $-z$ dock space with detectors at 35$^\circ$ and 215$^\circ$ in $\phi$. 

\subsubsection{VXD radiation-monitor and beam-abort system} 
\label{subsubsec:diamond_system}

A diamond-based detector system was designed to monitor the radiation dose rates in the interaction region, and to protect the VXD by triggering beam-abort requests in case of excessive beam losses, with a response time comparable to the revolution period of the accelerator beams.

\begin{figure}
	\begin{center}
		\includegraphics[width=.5\linewidth]{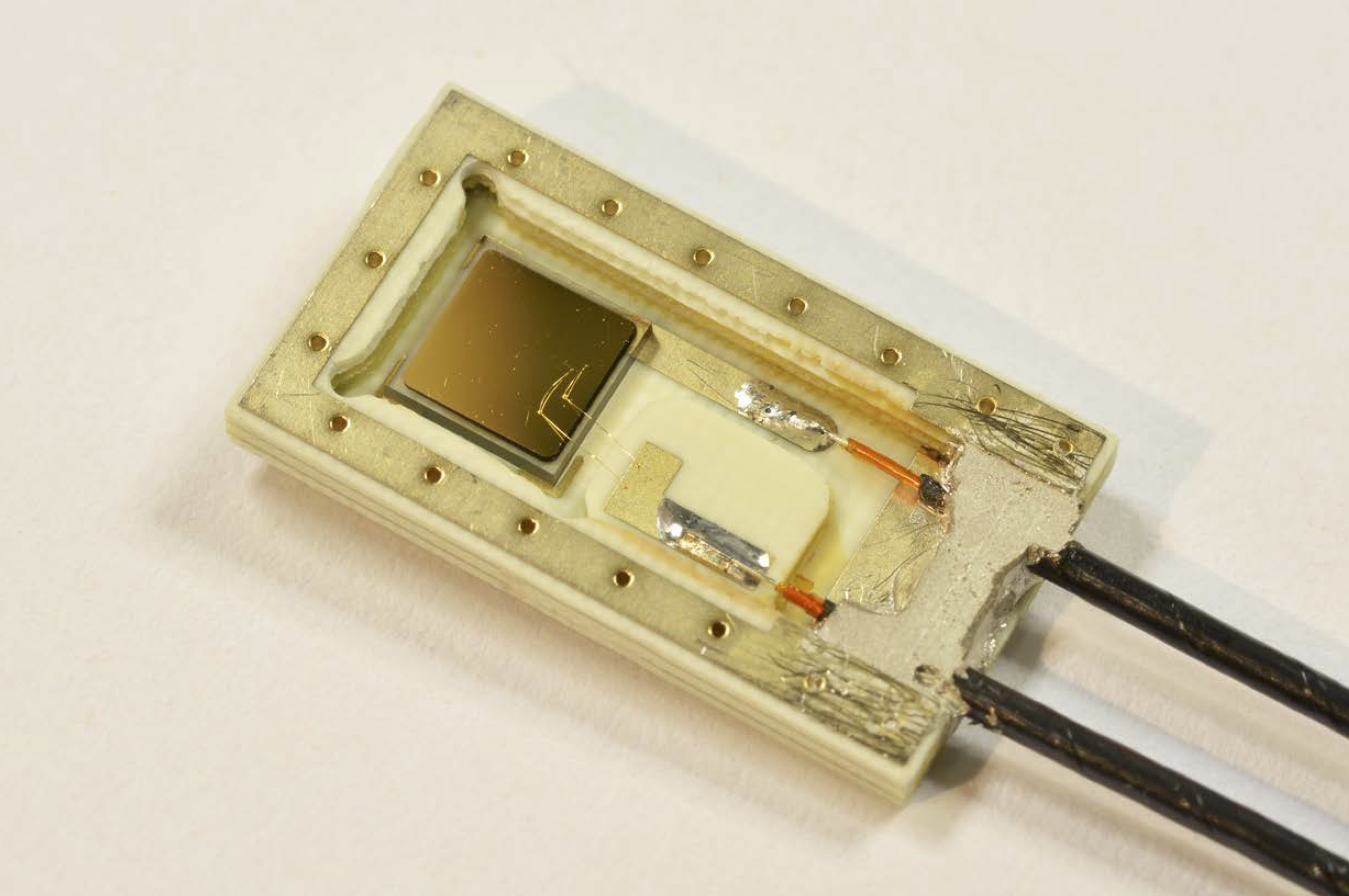} \\ 
		\includegraphics[width=\linewidth]{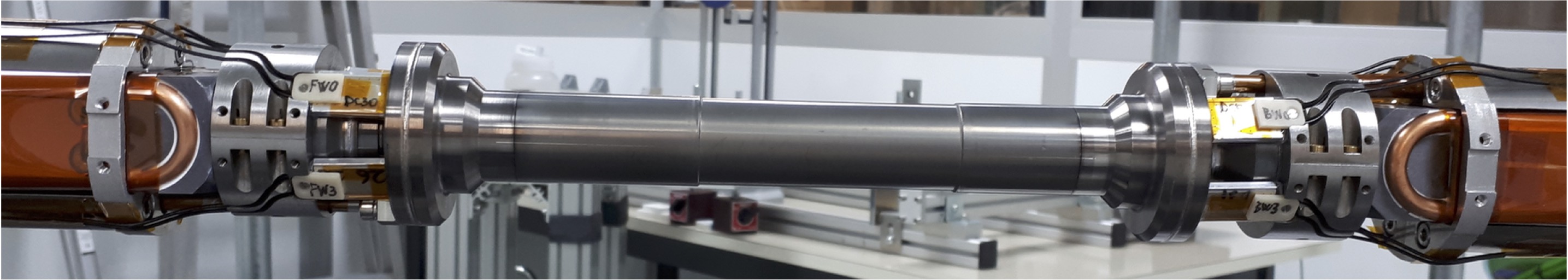} \\
		\includegraphics[width=\linewidth]{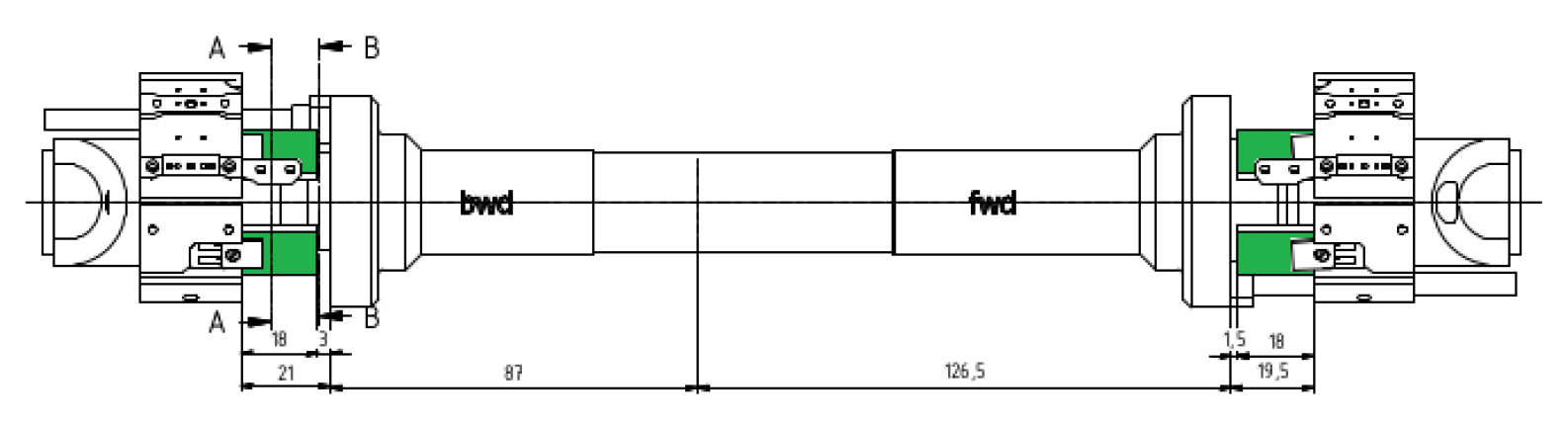}
	\caption{(top) Diamond sensor, used for radiation dose monitoring and beam abort triggering, mounted on its printed circuit board, without aluminum cover; (middle and bottom) diamond detectors mounted on the beam pipe: the diamond detectors are highlighted in green in the drawing.}
	\label{fig:diamonds_location}
	\end{center}
\end{figure}

For Phase~2, eight detectors were installed on the final version of the beam pipe, as shown in Figure~\ref{fig:diamonds_location}, forming a subset of the planned configuration of 28 diamond detectors for the VXD radiation monitoring and beam abort system of Phase~3. Their task is twofold: to contribute to the monitoring of beam losses, and to validate and tune the beam-abort function before the installation of the complete VXD in Phase~3.

After the initial dose-rate measurements at low beam currents, we performed validation tests of the abort thresholds and of the abort-signal exchanges with SuperKEKB in controlled conditions. Of the eight sensors, we devote four diamond detectors to monitoring, with the corresponding Diamond Control Unit (DCU) set to the lowest (most sensitive) current measurement range, and the remaining four detectors to the abort function, with the highest DCU current measurement range to measure large dose rates from the most severe beam losses.

We set the initial abort threshold at $1$\,rad, integrated in a moving time window of $1$\,ms (average dose rate of $1$\,krad/s in that time interval) chosen to protect the SVD silicon sensors, which could suffer localized damage from higher radiation levels.

The active part of each detector is an electronic-grade single-crystal $4.5 \times 4.5 \times 0.5$\,mm$^{3}$ diamond sensor, grown by the chemical vapor deposition technique, with two (Ti + Pt + Au) electrodes deposited on opposite faces. Each sensor is mounted on a small printed circuit board, providing mechanical support, electrical connections and screening, and is completed by a thin aluminum cover, as shown in Figure~\ref{fig:diamonds_location}.

The charge carriers, electrons and holes, produced by ionizing particles in the diamond bulk, drift in the electric field produced by the bias voltage applied to the electrodes, inducing a current proportional to the dose rate in the external circuit. Current-to-dose calibration factors were determined before installation with a pointlike $\beta$ source located at varying distances from the detector. For a measured current of $1$\,nA, the individually calibrated dose rates for the eight diamond sensors are in the $30 - 40$\,mrad/s range~\cite{Bassi:2021dno}.

The eight diamond detectors were controlled by two DCUs. The digital core of a DCU is a Field Programmable Gate Array (FPGA) that receives commands via an Ethernet interface, drives four independent high voltage modules, and accepts input data from an analog module with four input channels, including amplifiers and 16-bit analog-to-digital (ADC) converters with a $50$\,MHz sampling rate. 

By adding up the input data from the ADCs, the FPGA provides monitoring data at $10$\,Hz; the Experimental Physics and Industrial Control System (EPICS)~\cite{epics} read-out software applies pedestal subtraction and conversion to dose-rate units, to make these data available; the EPICS data are archived as process variables. One of three measurement ranges could be selected at initialization: $\pm 10$\,nA, $\pm 1$\,\si{\micro}A, and $\pm 1$\,mA. The corresponding sensitivity in $10$\,Hz data was about $0.3$\,pA, $30$\,pA, and $30$\,nA, repectively. The first range could detect dose rates down to a few \si{\micro}rad/s, while the third range could measure maximum dose rates of the order of $10$\,krad/s. 

An internal DCU revolving buffer-memory stores intermediate data at $100$\,kHz, with $10$\,\si{\micro}s time-granularity. Rolling sums of these data can be compared with abort thresholds at each $10$\,\si{\micro}s cycle, to provide beam-abort request signals for the HER and LER beams to the SuperKEKB beam-abort system. When activating the abort kicker magnets, SuperKEKB acknowledges the abort by sending back HER/LER abort timing signals. The DCUs use these signals to stop the writing of 100 kHz data into buffer memories; their frozen contents are read into files, which are then used for the analysis of the beam losses corresponding to each abort event.

\subsubsection{Luminosity} 

LumiBelle2, a fast monitor based on diamond detectors~\cite{Pang:2019ses}, recorded luminosity measurements. LumiBelle2 detects forward radiative Bhabha events at small ($<$\,\ang{0.1}) scattering angles and provides a measurement of the relative bunch-by-bunch luminosity at 1\,Hz, with precision up to 1\%, and of the relative integrated luminosity at 1\,kHz, with precision up to 0.1\%, depending on the luminosity. In addition, the Belle~II ECL also provides luminosity data.

\subsection{Online Real-time Feedback and Offline Integration} 

The decentralized nature of BEAST II and the need to unify the data for end-users, both for real-time monitoring and post-run analysis, required a dedicated infrastructure at both the hardware and software levels. To accommodate the large throughput of data from distributed sources, we use EPICS to distribute data in real-time to accelerator control, experimental operators for immediate feedback, and to a PostgreSQL database for later processing.

Several displays in the acccelerator control room  provide digested summary information from BEAST II detectors, Belle~II detectors, and the SuperKEKB accelerator. All observables for real-time monitoring were shared via EPICS and the displays themselves were constructed using Control System Studio \cite{css}.

The online displays provided visual markers to indicate undesirably high background levels for operation, but did not trigger an automatic abort. During the commissioning run, we added Belle~II subdetector information to the display, indicating a relevant quantity related to the observed background in each detector which could be correlated with the values from BEAST II. This real-time monitoring feedback system has since been expanded upon and remains in use in Phase~3.

Post-run (``offline'') data integration refers to the production of a single, common output file containing SuperKEKB accelerator conditions data, BEAST II detector background levels, and relevant quantities from individual Belle~II subdetectors. In order to provide a standard format for all of the information contained in this file, we record data at the 1\,Hz level and integrate over 1\,s intervals. We include all available values from BEAST II, Belle~II, and SuperKEKB, amounting to several hundred branches and thousands of channels. ROOT~\cite{ROOT} ntuples were provided on a day-by-day basis and also as shorter summaries of particularly relevant times, such as dedicated background runs or machine studies.


\section{Belle~II Operation Experience}
\label{sec:operational_experience}
\index{Belle~II Operation Experience}

Several major changes were implemented after the conclusion of Phase~1. While the most notable was the rolling-in of Belle~II, other changes included installation of QCS magnets around the IP, implementation of the new positron damping ring and installation of new collimators in the MR.

These changes resulted in a running environment substantially different from Phase~1. The presence of sensitive Belle~II detectors around the IP necessitated real-time measurement of background levels, to allow accelerator operators to adjust beam parameters to prevent excessive radiation doses and possible damage or detector performance degradation. 

Much of the Phase~2 beam time was dedicated to SuperKEKB accelerator machine tuning and studies as the accelerator team worked to reduce the vertical and horizontal beta functions at the IP, $\beta^{*}_y$ and $\beta^{*}_x$, and to improve accelerator performance, and later, to increase luminosity.

\begin{figure}
\begin{center}
\includegraphics[width=0.5\textwidth]{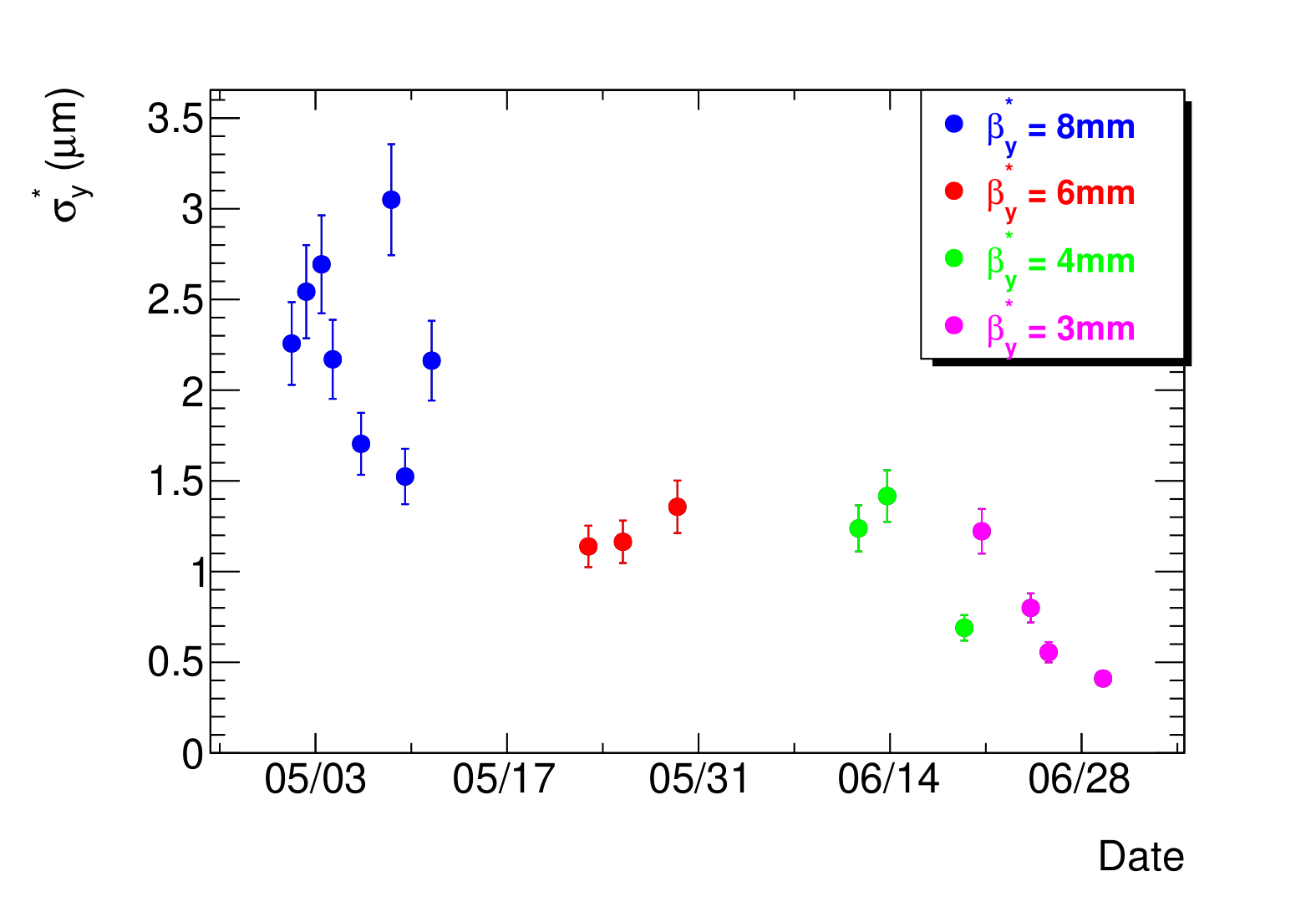}
\caption{Measured vertical beam size $\sigma_y$ at the interaction point during Phase~2 operation (2018), measured with the LumiBelle2 fast luminosity monitor~\citep{Pang:2019ses}.}
\label{fig:sigma_v_time}
\end{center}
\end{figure}

\subsection{Vacuum Baking}

A significant fraction of the SuperKEKB running time was devoted to vacuum baking to accelerate the process of outgassing from the beam pipe. During this time, beam was circulated without collisions in order to reduce the amount of residual contamination released during later data-taking runs. Belle~II was turned off during vacuum baking periods, but BEAST II always remained on.

\begin{figure}
\includegraphics[width=0.45\textwidth]{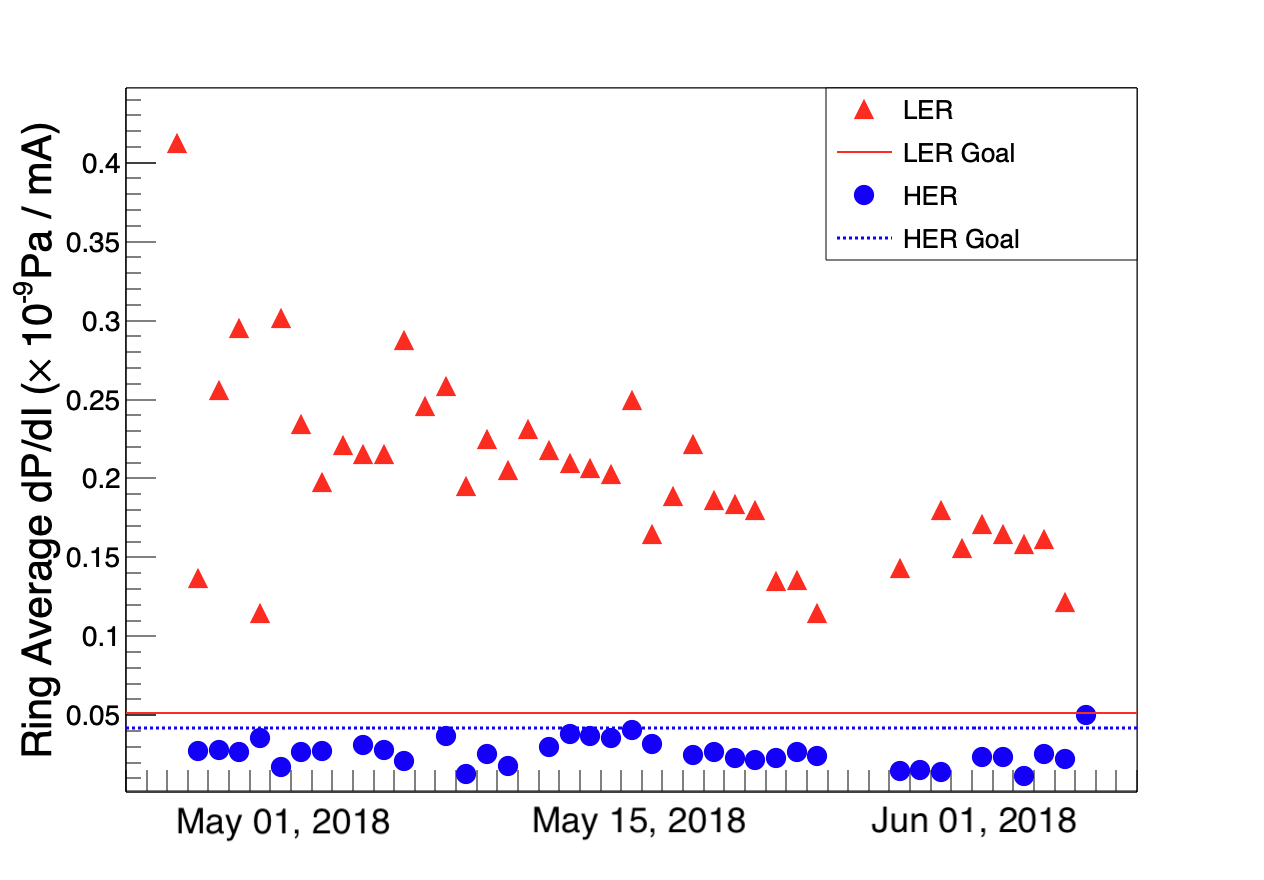}
\caption{Dynamic pressure in the LER and HER during Phase~2 running (2018). The LER shows a steady drop in gas pressure, while the well-conditioned HER has an order of magnitude lower dynamic pressure.}
\label{fig:dpdi}
\end{figure}

\begin{figure}
\includegraphics[width=.43\textwidth]{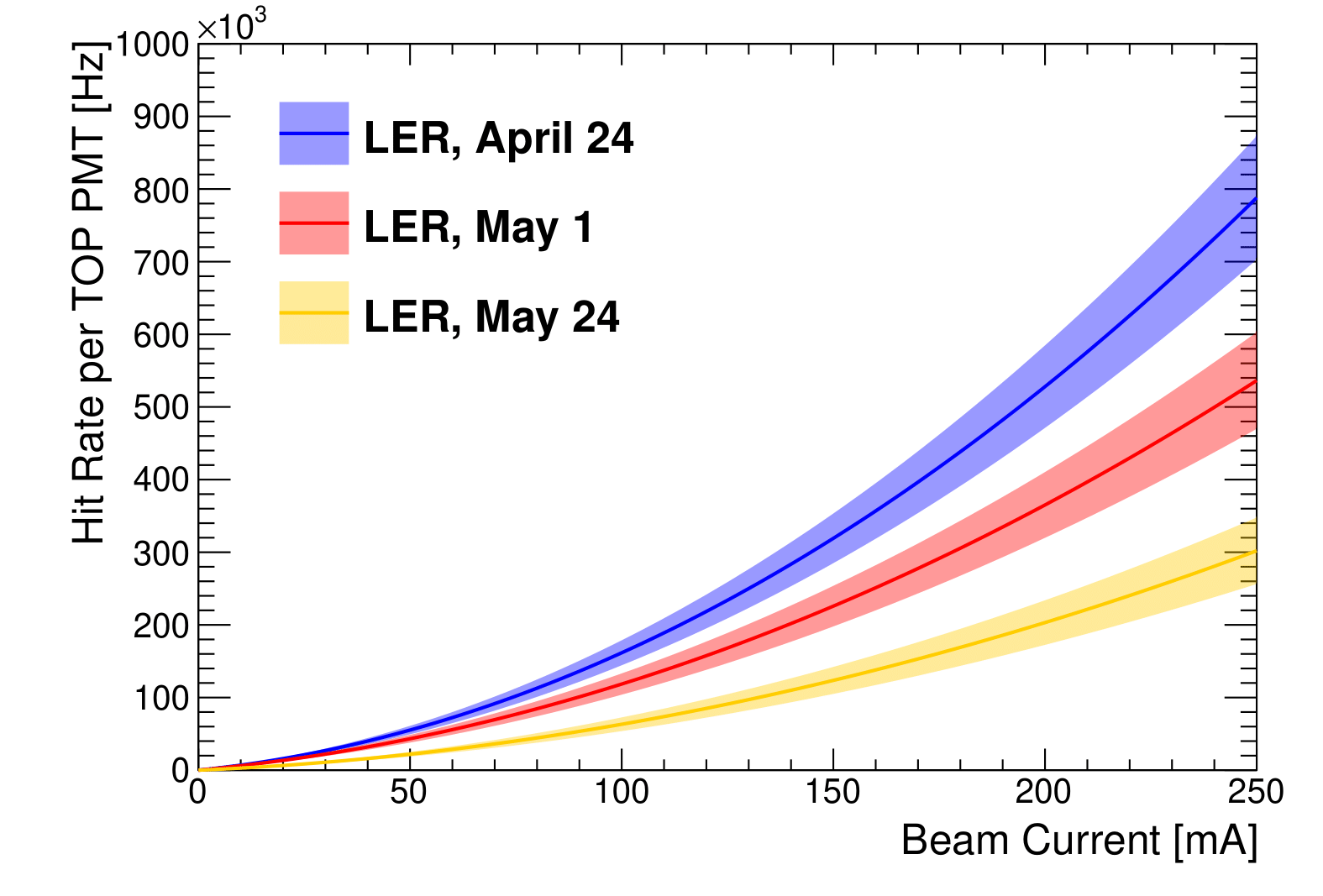}
\caption{(color online) LER backgrounds versus beam current, measured by the TOP during April and May 2018 demonstrating the effect of vacuum baking on background rates. From top to bottom, the lines correspond to beam doses of approximately 806\,A$\cdot$h,  829\,A$\cdot$h,  and 905\,A$\cdot$h.}
\label{fig:baking}
\end{figure}

In particular, the LER beam pipe was replaced between Phases 1 and 2 and contained considerably more residual gas than the older HER, consistently resulting in higher beam pressure levels and therefore larger beam-gas backgrounds. The dynamic pressure (slope of pressure versus current) in both the LER and HER are shown in Figure~\ref{fig:dpdi} and an example of the effect on detector backgrounds is presented in Figure~\ref{fig:baking}.

\subsection{First Collisions and Luminosity}

The addition of the QCS magnets and final focusing system allowed SuperKEKB to focus and collide beams. In the early hours of April 26${\mathrm{th}}$, 2018, slightly more than one month after beams first started circulating, Belle~II recorded its first electron-positron collisions. As Phase~2 progressed, the luminosity delivered to Belle~II steadily increased as a result of shrinking beam sizes and increased currents. By the end of Phase~2, SuperKEKB delivered a peak luminosity of $5.6 \times 10^{33} \text{cm}^{-2} \cdot \text{s}^{-1}$ and Belle~II had collected a total integrated luminosity of $472\,\mathrm{pb}^{-1}$.

\subsection{Creation of the BCG Group}

To prevent excessive radiation dose on the Belle~II subdetectors, close communication with the SuperKEKB accelerator team and Belle~II detector operators is necessary. To meet this need, members of Belle~II and SuperKEKB together created a new Belle~II Commissioning Group (BCG). BCG members served as liaisons between accelerator and detector operators. The introduction of real-time background displays as well as the presence of a BCG operator in the accelerator control room meant feedback could be immediately exchanged between accelerator and detector operators to ensure smooth operation and favorable beam conditions for Belle~II data collection.

\subsection{Dedicated Background Studies}

The BCG also designed and carried out dedicated background studies during the course of Phase~2. In general, these studies measured the background composition and effect of a particular change in beam parameters on the background rates in both BEAST II and Belle~II detectors. Examples of study parameters include beam size and emittance, beam steering angle, horizontal and vertical collimator aperture widths, and interaction vertex position. 

As beam parameters were varied, the effect on background rates became immediately visible in the online displays, allowing for determination of optimal running conditions.

\subsection{Magnet Quenches}

A consistent issue during operation of SuperKEKB in Phase~2 was quenches occurring in the superconducting steering magnets of the final focus. During the period from April 1st to May 17th, 2018, a total of 23 magnet quenches occurred, each of which was accompanied by a beam abort and followed by an investigation. Quenches occurred in numerous areas around both the LER and HER.

We identify causes from a variety of sources, including injection kicker instabilities, beam parameter changes, machine tests, and other operational issues. Most quenches occurred during beam injections, but 6 of the 23 quenches occurred during beam storage. Because the sudden loss of a magnet affects beam trajectories unpredictably, damage to both sensitive detector components as well as the QCS magnet system near the IP is a serious concern; in addition, the recovery from a quench implies the loss of significant operation time. Therefore, we devoted a great deal of effort to reduce the frequency of such events. 

For 19 quench events, the data from the DCU buffer memories with 10\,\si{\micro}s time resolution in 1\,s time windows allowed a detailed study of the beam-loss pattern preceding the quench event in the interaction region. In 15 out of 19 events a lower diamond abort-threshold would have triggered a beam abort, preventing the magnet quench. 

As a result of this study, we redefined the abort thresholds: the ``fast" abort condition, aimed at detecting rapidly increasing radiation doses, kept the moving integration time-window of 1\,ms, but lowered the integral threshold value from 1\,rad to 10\,mrad; we also introduced a new ``slow" abort condition developed to avoid the long-term accumulation of slowly increasing radiation doses. For the latter signal, we set the corresponding threshold to a 200\,mrad dose integrated in a 1\,s moving time window. In both cases, a signal above threshold from at least two diamond detectors simultaneously generated an abort trigger.

\begin{figure}[htb]
    \begin{center}

  \includegraphics[trim=0 0 0 85, clip, width=.45\textwidth]{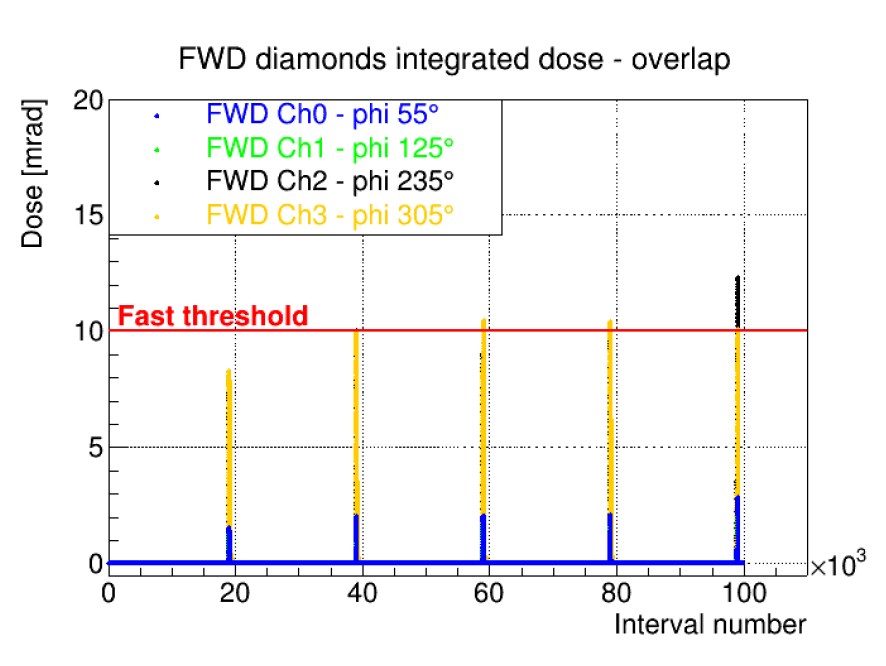}
    \includegraphics[trim=0 0 0 85,clip,width=.45\textwidth]{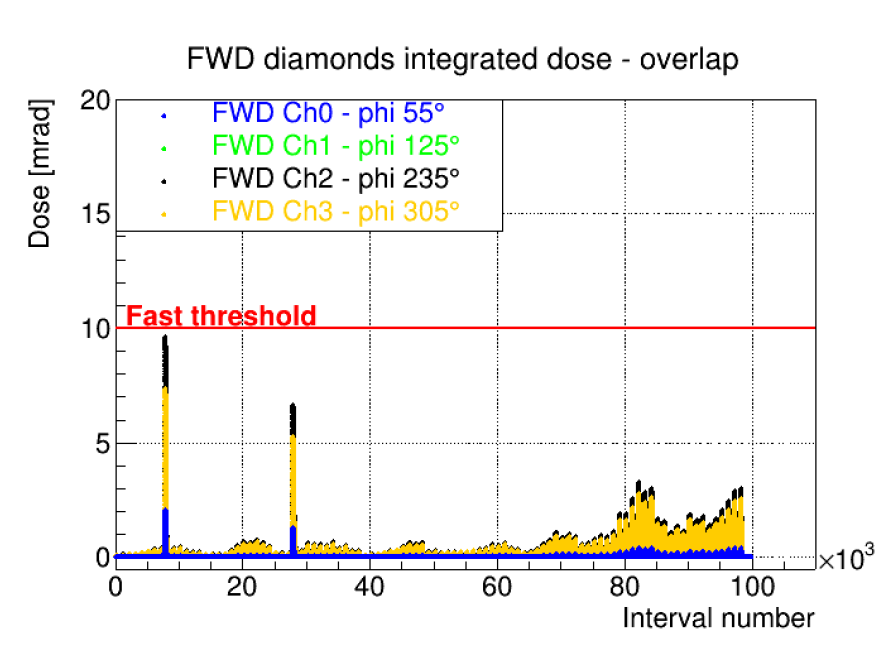}
    \caption{Moving sums of radiation doses measured by four diamond sensors, immediately preceding two aborts. The top (bottom) panel shows an example of a fast (slow) abort. Details are provided in the text. The horizontal axis refers to the 100 $\times$ 10$^3$ moving time windows in the 1\,s time interval.}
  \label{fig:abort_example}
  \end{center}
\end{figure}

This new configuration proved to be effective in preventing QCS quenches. Figure~\ref{fig:abort_example} shows two examples of moving sums of dose rates over 1\,ms intervals, reconstructed from the data stored at 100\,kHz, every 10\,\si{\micro}s, in the DCU buffer memories. The \emph{interval numbers} on the abscissa correspond to the 100\,$\times 10^{3}$ moving time windows in a 1\,s time interval. In the top panel, beam losses are small except during the injection at 5\,Hz, causing the five evident peaks. At the last peak, two moving sums exceed the ``fast"  threshold and trigger the abort. In the bottom panel, the losses correlated with injection are not sufficient to trigger a ``fast" abort, but an oscillating pattern of slowly increasing losses triggers the ``slow" abort.

\subsection{Background Storms}

On several occasions in Phase~2, we observed unusually high levels of background in multiple Belle~II detectors for short periods of time. These were chiefly evident in the PXD and SVD, but also appeared in detectors farther from the IP such as the TOP. 

\begin{figure}[h]
  \begin{center}
  \includegraphics[width=.45\textwidth]{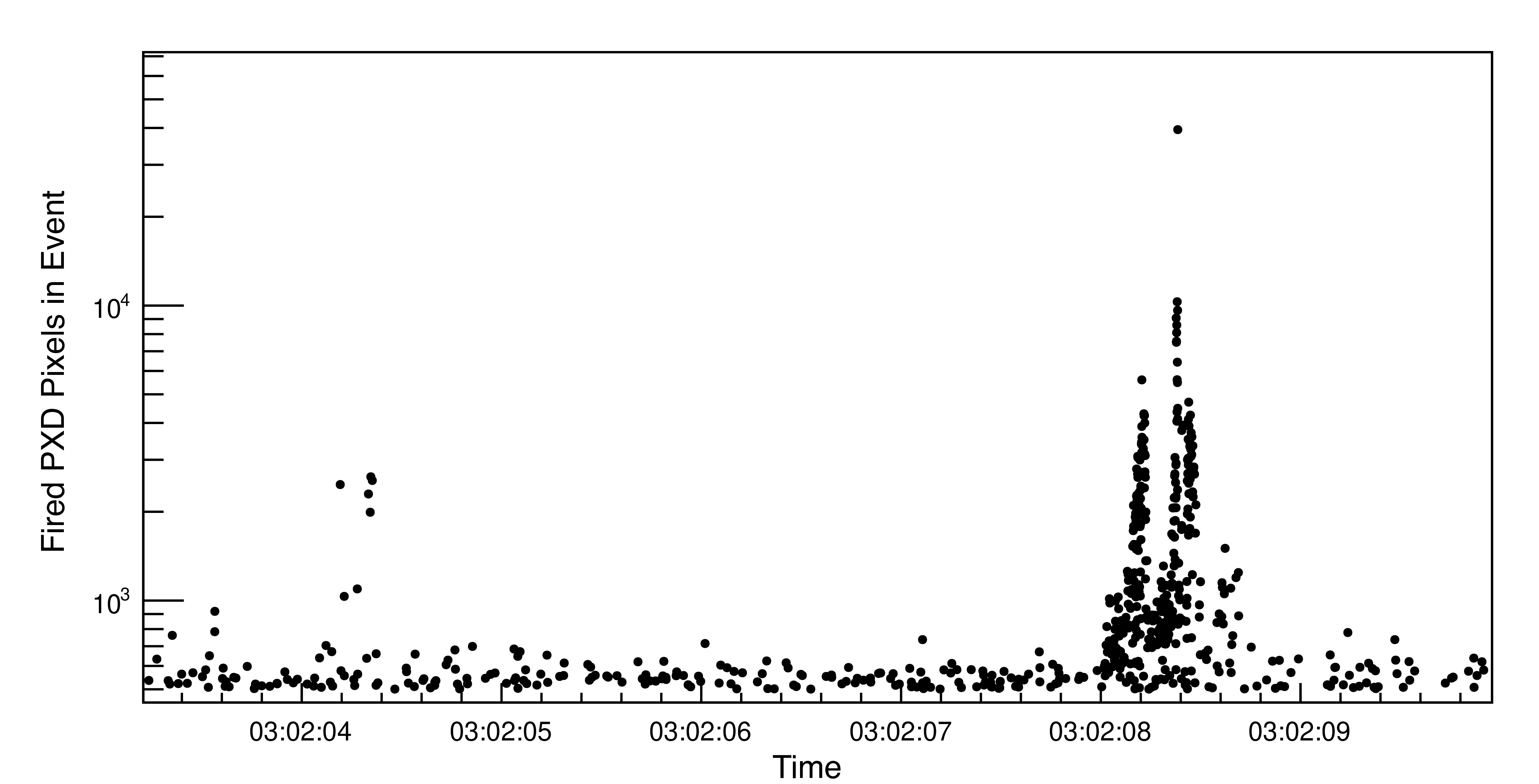}
  \caption{Example of a background storm event recorded in the PXD during Belle~II operation. A sudden rise in detector activity an order of magnitude above base levels is briefly visible before returning to normal.}    
  \label{fig:bg_storm}
  \end{center}
\end{figure}

\begin{figure}[h]
  \includegraphics[trim=0 0 0 100,clip,width=.50\textwidth]{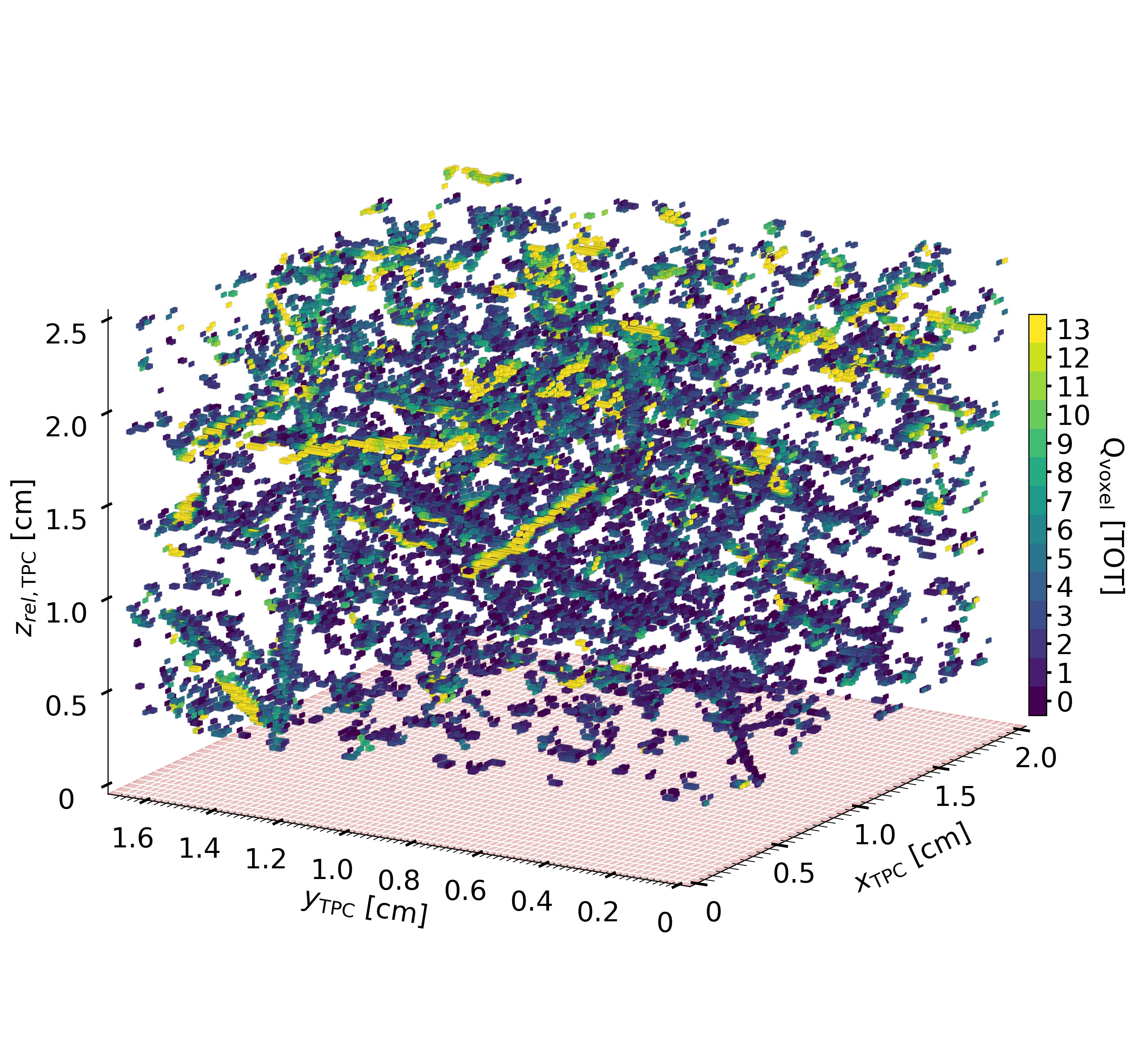}
  \caption{(color online) Example of ionization observed in a TPC fast neutron detector during a background storm event, integrated over a period of 0.056\,s. The colored tracks and clusters represent individual ionization events, mostly from neutral particles. The thicker tracks are nuclear recoils initiated by fast neutrons, the smaller clusters are x-ray conversions. For comparison, during normal conditions we observe neutrons at a rate of order 1\,Hz.}
  \label{fig:n_storm}

\end{figure}

Despite efforts to correlate these events, referred to as ``background storms", with accelerator conditions and particularly with magnet quenches, we find no correlation. Background storms occured in both beam storage and injection periods. An example of data recorded during a background storm is shown in Figures \ref{fig:bg_storm} and \ref{fig:n_storm}, as recorded in the PXD and TPC systems, respectively. In the latter case, we estimate that the neutron flux at the TPCs was as high as $6 \times 10^{5}$\,Hz\,cm$^{-2}$, approximately three orders of magnitude above normal levels.

\subsection{Beam Steering Corrections}

During the course of Phase~2, a persistent background contribution incompatible with the model discussed later in Section \ref{sec:heuristic} appeared in the LER. Hypothesized origins of this background include unexpectedly long horizontal beam tails, beam scraping on the IR beam pipe walls and the resulting showers, and collimator effects.

Attempts to lessen the effects of beam scraping backgrounds included beam orbit and angle tuning, along with adjustment of the collision vertex position. We determined that the HER angle at the IP was sub-optimal, and adjustment led to rapid improvements in beam conditions. 

In addition, analysis reveals that the longitudinal position of the collision vertex was initially several mm away from the nominal position, causing unacceptably large numbers of tracks with origins away from the intended IP. Using iterative LER steering corrections, the real collision vertex was brought into line with the nominal position. 
\section{Background Simulation}
\label{sec:simulation}
\index{Background Simulation}

The beam background simulation methodology used here is the same as in Phase~1~\cite{Lewis:2018ayu}. The main difference is in the improvement of the simulated geometry. Simulation of synchrotron radiation is a separate effort, described in Section~\ref{sec:synchrotron_sim}.

\subsection{Single-beam backgrounds: Coulomb, Bremsstrahlung, Touschek} 

Touschek and beam-gas scattering (Coulomb and Bremsstrahlung) result in beam particles that deviate from the nominal orbit. While the initial scattering can occur at any location around the ring, scattered particles tend to stop at specific locations. In particular, they are most frequently stopped by beam collimators or hit the beam pipe inside the QCS magnets nearest the interaction point where the physical aperture is narrowest with respect to the beam size.

Showers generated inside the QCS can result in secondary particles reaching Belle~II. In order to simulate these types of background accurately, a large number of components must be modeled properly: the initial scattering probability, beam optics, collimators, beam pipe shape inside the QCS, shielding material, and the Belle~II detector. We simulate the scattering, the optics, and the collimators in the Strategic Accelerator Design (SAD) beam particle tracking code~\cite{SADHP}.

SAD includes simulation of Touschek, beam-gas scattering (both Coulomb and Bremsstrahlung processes), while collimator tip scattering and injection background simulations are not implemented. Subsequent to the work presented here, SAD was embedded in a larger software framework that enables tip scattering simulation \cite{natochii2021improved}, but Phase~2 simulations with tip scattering do not exist.

The simulation gives the beam particle loss rates in the whole ring and in the interaction region, defined as $ \pm 4$\,m from the IP, for each background component. Losses in the interaction region are used as a proxy for Belle~II background rates in collimator optimization simulations, where full Geant4~\cite{AGOSTINELLI2003250} simulations would be probitive due to CPU requirements.

We consider a safe limit for background in the interaction region (IR) to be 100\,MHz including both HER and LER contributions.

For more precise Belle~II background simulations, Geant4 is used in addition to SAD. The beam pipe shape is implemented both in SAD and Geant4. If a beam particle crosses the beam pipe envelope in SAD, it is passed to Geant4 and the subsequent showering and Belle~II digitization is simulated with the core Belle~II software, basf2~\cite{Kuhr_2018,the_belle_ii_collaboration_2021_5574116}.

\subsection{Luminosity Backgrounds}

Undesirable backgrounds in the Belle~II detectors due to colliding beams are substantially easier to simulate. 
Such backgrounds are produced at the IP and for that reason are largely independent of beam optics and collimator settings. 
A detailed tracking of off-orbit particles around the ring is not required in this case. 
We simulate luminosity backgrounds in the same way as Belle~II physics, with designated event generators followed by Geant4.

\subsection{Synchrotron Radiation} 
\label{sec:synchrotron_sim}

The generation of primary charged particles and simulation of {\SR} photon emission during propagation through the magnetic field requires a very large amount of CPU time to get a complete picture of the {\SR} background in detectors. For this reason, we simulate SR photons using vertex and momentum information from the primary particle simulation, which performs tracking of the beam particles through the beamline.
We perform both primary and final simulations with Geant4 within the basf2 framework.
A detailed description of {\SR} background generation is described in Reference~\cite{Lewis:2018ayu}.
For clarity, we present {\SR} simulation results together with the measurements in Section~\ref{sec:sr_results}.


\section{Experimental Results}
\label{sec:exp_results}

\subsection{Beam-Gas and Touschek Study}
\label{sec:bg_tous_study}
\index{Beam-Gas and Touschek Study}

	In the final two months of Phase~2, we carried out four major beam background studies. To avoid contamination from backgrounds arising from beam-beam interactions, each study was carried out with only a single beam (HER or LER)  active. These studies are summarized in Table \ref{table_bg_studies}.
		
	\begin{centering}	
\begin{table}[h]

\begin{tabular}{| l | c | c | c | }
\hline
Date & Study Type & $\beta_y^*$ (mm) & $\beta_x^*$  (mm)\\ \hline
June 11 & HER Single Beam &  4  & 100\\
June 12 & LER Single Beam  & 4 & 100\\ 
July 16 & HER/LER Beam Size & 3 & 100 \\ \hline

\end{tabular}
\caption{Phase~2 single-beam background studies. The July 16 study was two individual single-beam studies carried out sequentially.}

\label{table_bg_studies}
\end{table}%
\end{centering}

\begin{figure*}[h]
  \begin{center}
    \includegraphics[width=\textwidth]{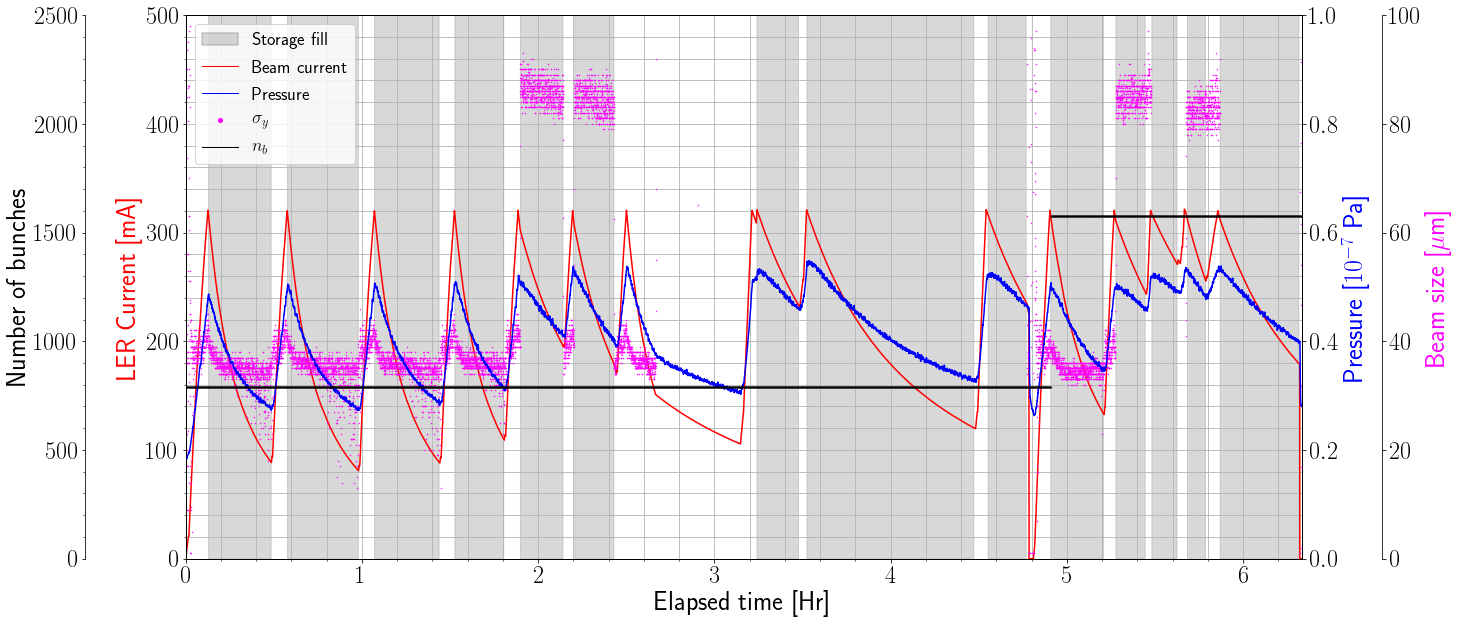}
    \caption{(color online) Accelerator data recorded during the LER study performed on June 12th. Data within the shaded regions correspond to beam storage fills that were included in background analyses. Vertical axis labels are colored to match the color of their corresponding plotted observables. Beam size values greater than \SI{100}{\micro\meter} are omitted.. All data are from the LER. }
    \label{fig:storage}
  \end{center}
\end{figure*}

        We use only data taken during beam storage periods in our analysis to avoid the higher background levels present during beam injections. \autoref{fig:storage} shows a snapshot of relevant accelerator data during a background study, with selected data shown in shaded areas.  For each of the four studies, we fit the combined background model to experimental data in an attempt to determine the contribution of each source to the overall background level. Fits were performed individually for each BEAST II and Belle~II detector.
        
\subsubsection{Measured Gas Composition}

We measure gas pressures around the ring with 300 monitors positioned in individual ring sections, as described in Section \ref{sec:Beam_Gas_Pressure}. To provide a more realistic summary of the gas conditions in the beamline, an empirical calculation to account for contributions from individual gas species is carried out using data from three RGAs instrumented along the SuperKEKB beamline. Each RGA is a mass spectrometer providing partial pressures of m/$Z$ values from 1 to 50. In principle, each m/$Z$ represents a single species of molecule in the beam-gas mix, although certain species overlap.
 
 Using RGA data, we calculate a single effective $Z$ value for all species in the gas, denoted $Z_{\text{eff}}$, which is used when fitting the data, as described in Section~\ref{sec:heuristic}. For the period considered in Phase~2, the calculated $Z_{\text{eff}}$ consistently falls between 2 and 2.5. 
In contrast, the simulation assumes a time-independent, uniform value of $Z_{\text{eff}} = 7$ around the ring.
 
\subsubsection{Analysis Methodology} 
\label{sec:heuristic}

Based on the relationships between background types and detector rates described in Section \ref{sec:types}, we posit a combined background model to describe the contributions of Touschek and beam-gas events to the observed background rate $R$:

\begin{center}

\begin{equation}
R = B \cdot I \cdot P Z_{\text{eff}}^2 + T \cdot \frac{I^2}{\sigma_y  n_\text{b}}
\label{eq:heuristic}
\end{equation}
\end{center}
where $B$ and $T$ represent the beam-gas and Touschek rates. It is convenient to rearrange Eq.~\ref{eq:heuristic} to allow for a graphical representation:

\begin{equation}
\label{eq:heur_linear}
\frac{R}{I \cdot P Z_{\text{eff}}^2} = B + T \cdot \frac{I}{PZ_{\text{eff}}^2\sigma_y n_\text{b}}
\end{equation}

If the left hand side of Eq. \ref{eq:heur_linear} is plotted versus $\frac{I}{P Z_{\text{eff}}^2 \sigma_y n_\text{b}}$, we expect a line with a slope proportional to the Touschek sensitivity, T, and an offset corresponding to the beam-gas sensitivity, B.

\subsubsection{Measurements of Background Composition}

\begin{figure}[h]
\begin{center}
\includegraphics[width=0.45\textwidth]{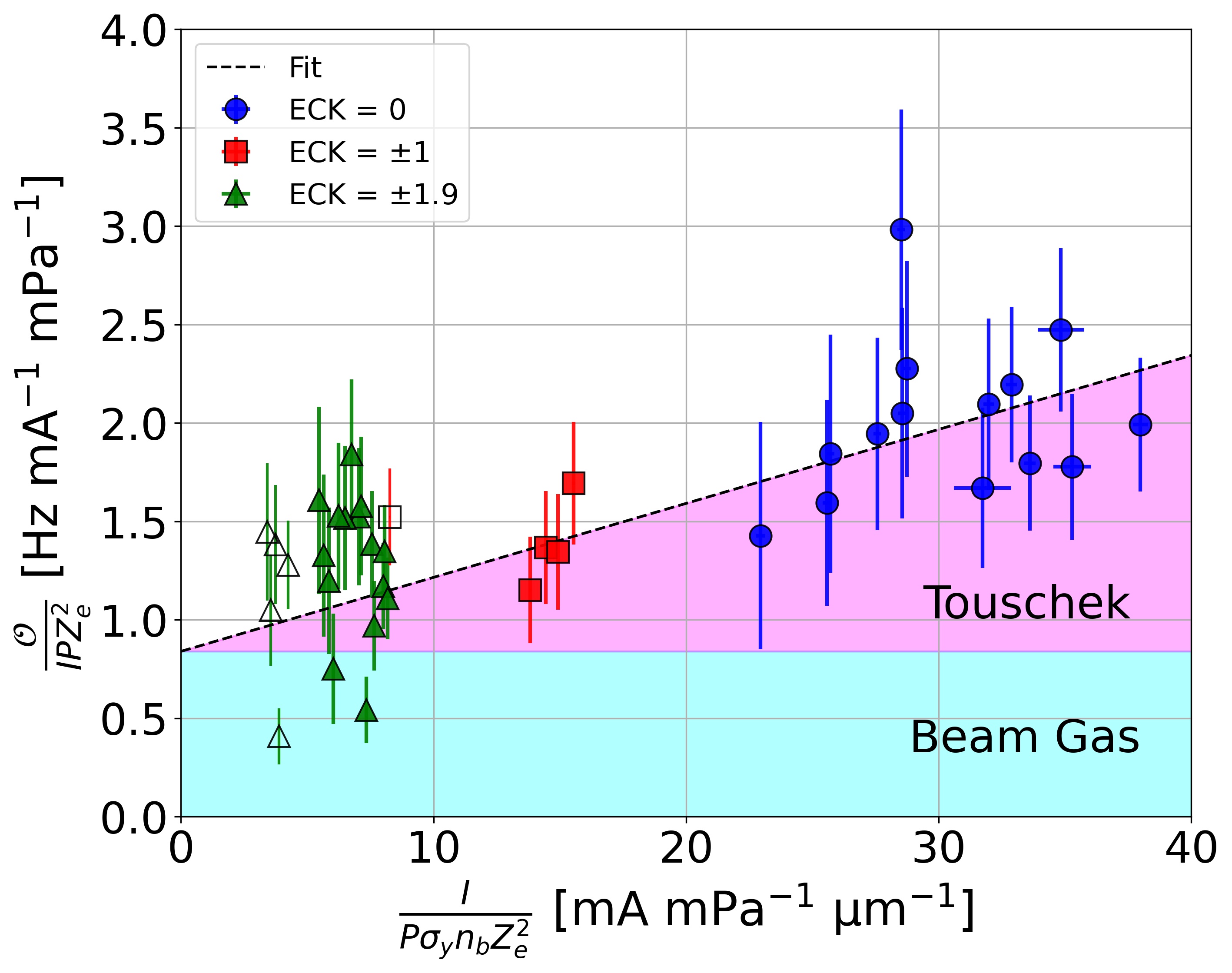}
\caption{Example of using Eq. \ref{eq:heuristic} to separate beam-gas and Touschek contributions to observed background rates, for a TPC during the June 12th LER study. Filled points represent $n_\text{b}=789$, and unfilled points represent $n_\text{b}=1576$. ``ECK" (legend) refers to the setting of the emittance control knob used to tune the vertical beam size. Larger magnitudes of the knob setting value correspond to larger $\sigma_y$. Touschek contributions increase relative to beam-gas contributions with decreasing beam size.}
\label{fig:heuristic}
\end{center}
\end{figure}

\begin{figure}[h]
\begin{center}
  \includegraphics[width=0.5\textwidth]{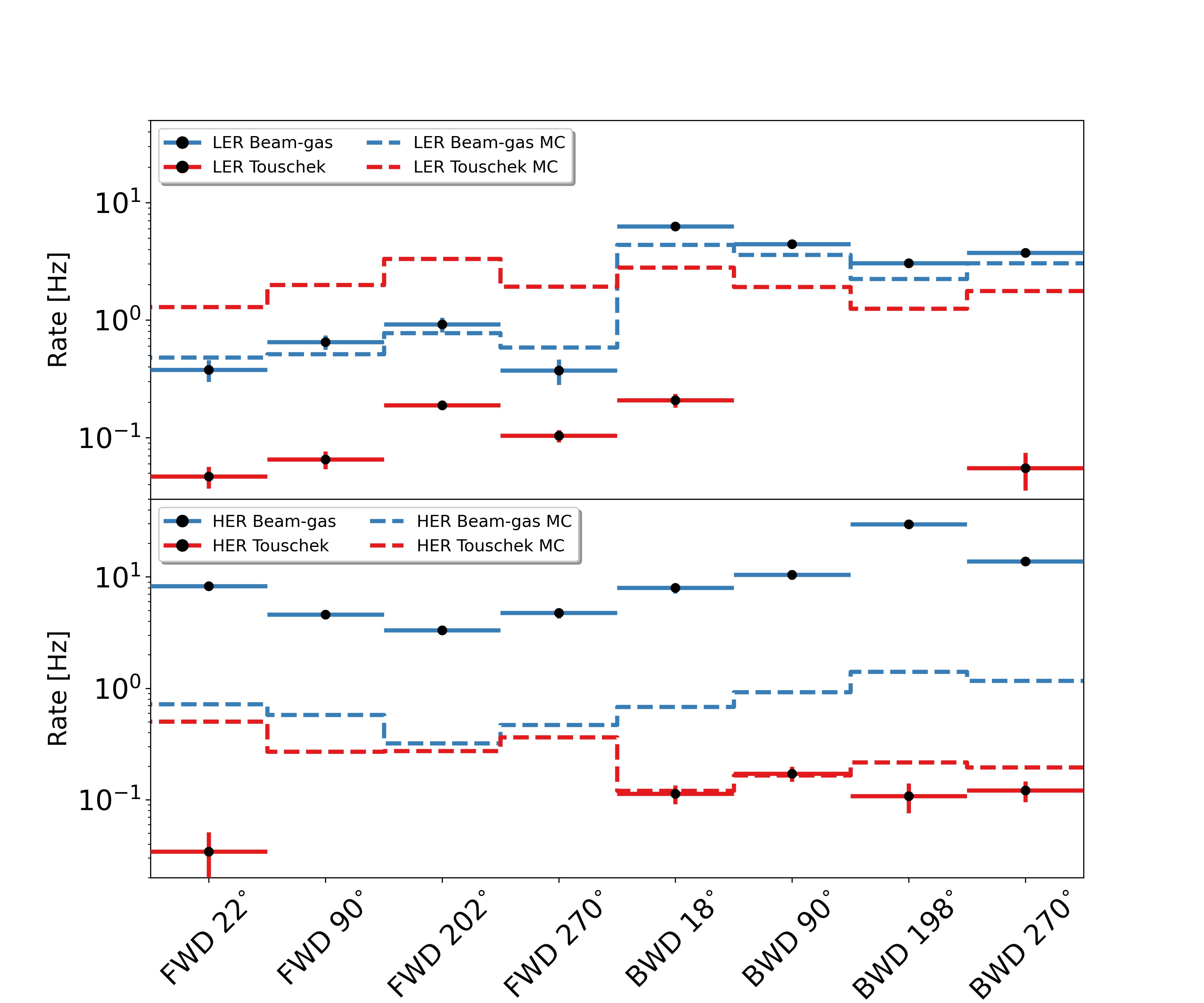}
  \caption{Summary distributions of nuclear recoil rates observed in all TPCs during the LER (top) and HER (bottom) studies with $\beta^*_y=4$\,mm optics. The TPCs are labeled by their absolute $\phi$ positions in degrees and their relative $z$ positions (BWD and FWD represent $z = -\SI{1.3}{m}$ and $z = +\SI{1.9}{m}$, respectively). Solid lines correspond to experimental data and dashed lines correspond to MC. Experimental rates are extrapolated to the accelerator conditions used in MC using the LER and HER beam-gas and Touschek fit coefficients extracted from combined heuristic fits to measured data. Beam-gas scattering dominates neutron production, and the spatial distribution of neutrons from that process appears to be accurate in the simulation. Touschek contributions not plotted are beneath the lower limits of the plot.}
\label{fig:tpc_phi}
\end{center}
\end{figure}

We fit background rates observed in individual detectors to the linear background model thus described by Eq.~\ref{eq:heur_linear}, from which we obtain relative contributions of background sources. An example fit for one of the TPC neutron detectors is shown in Figure \ref{fig:heuristic}. We see clearly how varying the beam size allows us to separate observed neutron rate contributions from beam-gas and Touschek scattering. Figure~\ref{fig:tpc_phi} shows how the resulting measured background composition in each of the eight TPCs (located at four $\phi$ locations on either side of the IR) compares against the background simulation. We see that the spatial distribution of neutrons from beam-gas scattering is remarkably accurate in the simulation, for both rings. In experiment, this background is the leading source of neutrons. The beam-gas normalization is also well predicted for LER. For HER, however, observed beam-gas background exceeds predictions by one order of magnitude.

\subsubsection{Evaluation of Simulation Accuracy}
\label{sec:ratio_summary}

\begin{figure}[h]
\includegraphics[width=0.45\textwidth]{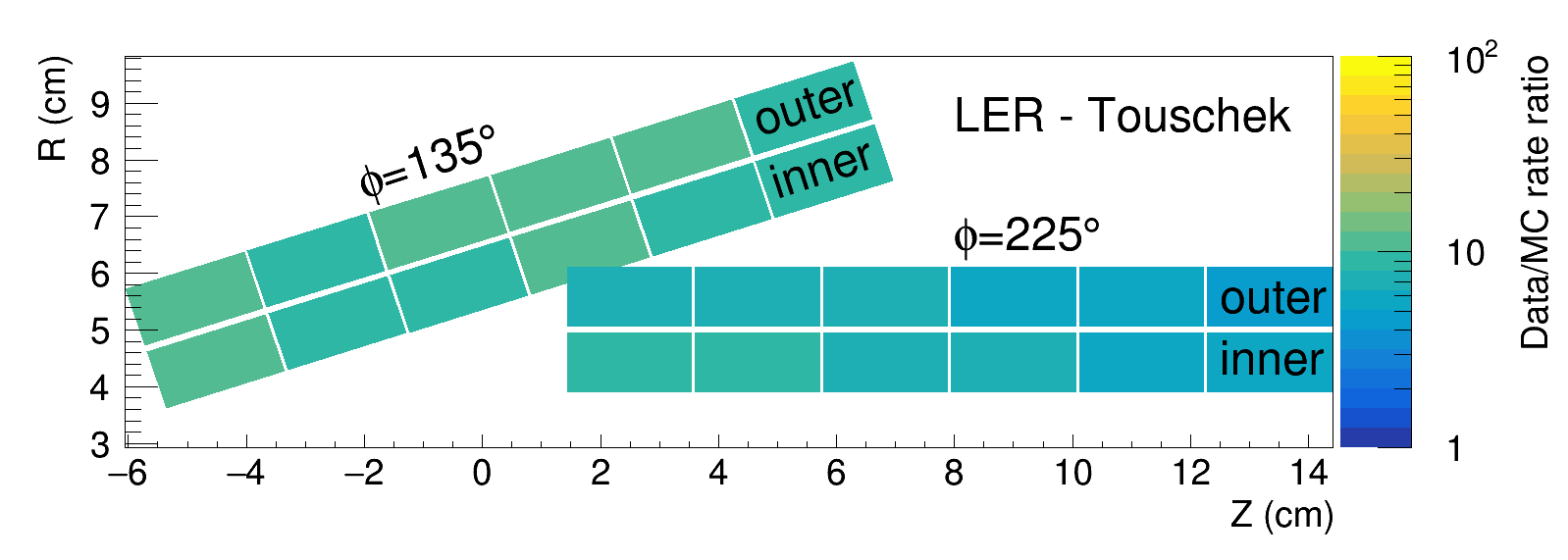}
\includegraphics[width=0.45\textwidth]{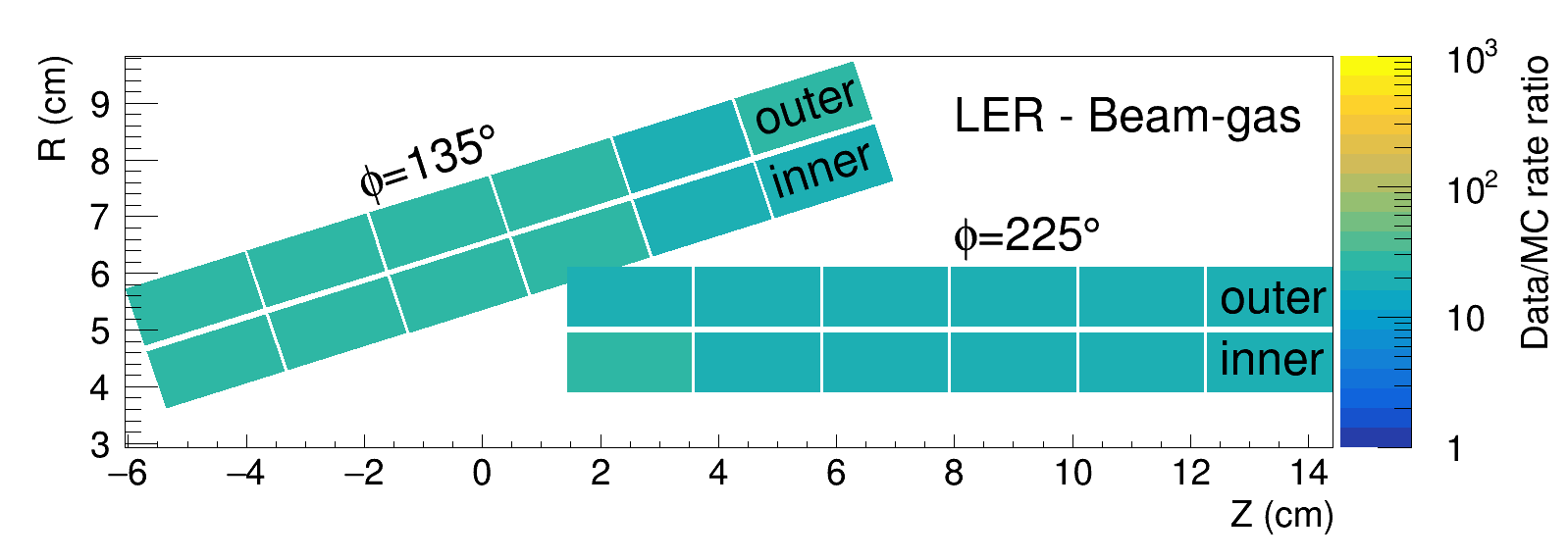}
\caption{Background data/MC ratios as measured in PLUME during the June 12th LER study. Ratios are shown separate for Touschek (top) and beam-gas (bottom). Both background components are substantially larger than predicted by simulation, with the excess having no strong position dependence.}
\label{fig:plume_ratio}
\end{figure}

We evaluate the accuracy of our simulation and its consequent predictive ability by comparing predicted detector rates in all BEAST II and Belle~II detectors to those measured during dedicated background studies. Touschek- and beam-gas induced rates are calculated separately as described in Section~\ref{sec:heuristic} and divided by the simulated prediction to produce a ratio for each detector channel. In the ideal case, a ratio of 1 indicates agreement between measurement and simulation. For the detectors with high hit rates, the fitting procedure and comparison can be performed even for individual channels. This provides highly granular, position-dependent information on the simulation accuracy for individual background components, as shown in Fig.~\ref{fig:plume_ratio} for PLUME. 

\begin{figure}[h!]
\begin{center}

\includegraphics[trim=0 0 60 0,clip,width=0.45\textwidth]{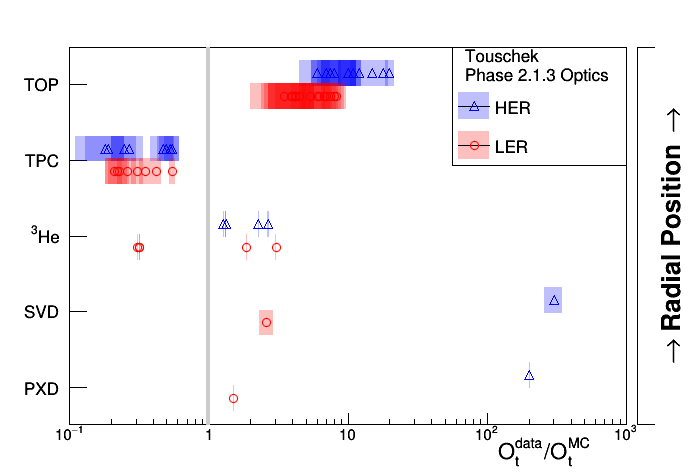} \\
\includegraphics[trim=0 0 60 0, clip,width=0.45\textwidth]{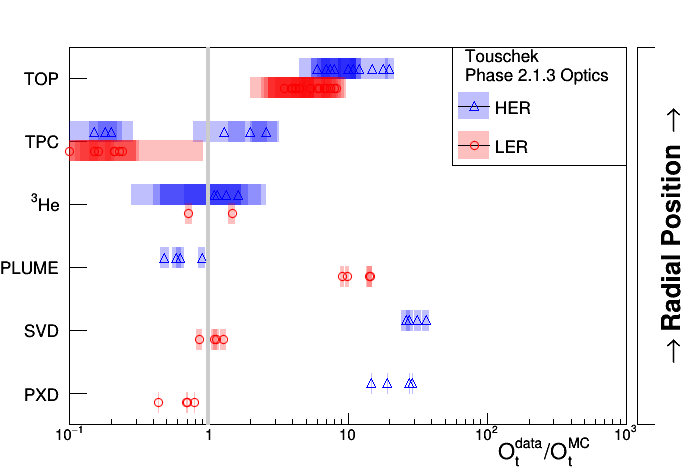}
\caption{(color online) Ratio of observed to predicted Touschek background rates in all detectors studied with old (top) and new (bottom) simulation.  Blue (Red) points represent HER (LER) results. From top to bottom, the detectors are ordered from radially outermost (TOP) to inermost (PXD).}

\label{fig:tous_summary}
\end{center}
\end{figure}

\begin{figure}[h!]
\begin{center}
 \includegraphics[trim=0 0 63 0, clip,width=0.45\textwidth]{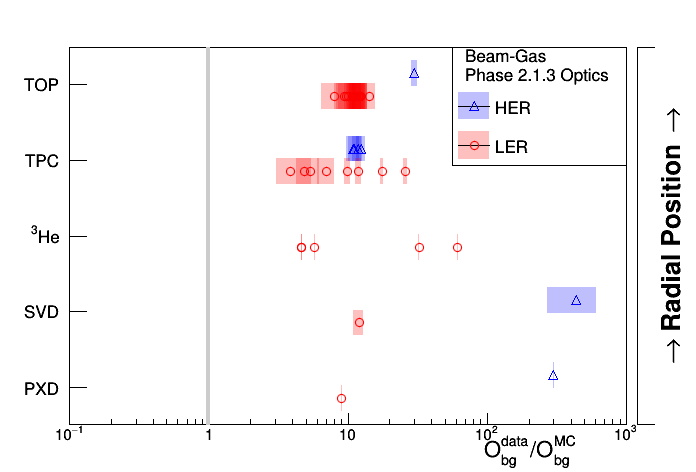} \\
 \includegraphics[trim=0 0 60 0, clip,width=0.45\textwidth]{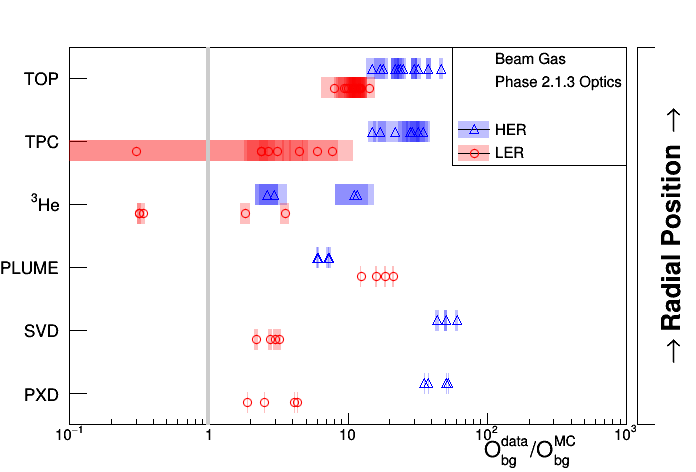}
\caption{(color online) Ratio of observed to predicted Touschek background rates in all detectors studied with old (top) and new (bottom) simulation.  Blue (Red) points represent HER (LER) results. From top to bottom, the detectors are ordered from radially outermost (TOP) to inermost (PXD).}

\label{fig:bg_summary}
\end{center}
\end{figure}

Data-to-simulation Monte Carlo (Data/MC) fit results for all BEAST II and Belle~II detectors are summarized in~\autoref{fig:tous_summary} for Touschek backgrounds and~\autoref{fig:bg_summary} for beam-gas results. Values for individual detector channels or physical locations, where applicable, are shown as separate points. The top plot in each figure represents the results of the Data/MC fits using the ``old" simulation, while bottom plots show the same results using MC updated to better model the detector. Data/MC fit results are improved markedly with the new simulation, usually by orders of magnitude. Table~\ref{table:avg_of_avg} combines the individual detector results into a single overall ratio for Touschek and beam-gas backgrounds in the LER and HER. In Section~\ref{sec:summary_recommendations} we use these data to extrapolate to expected beam conditions in Phase~3.

\begin{table*}
\begin{center}
 \begin{tabular}{ | c | c | c | c | c | c | }
\hline
Ring & Background Source & October 2018 Simulation & February 2019 Simulation & October 2018/February 2019 Ratio \\ 
\hline
\multirow{2}{*}{HER} & Touschek & 127.82  &  113.91 & 1.12  \\   
 &   Beam-gas & 483.50 & 32.28 & 14.98  \\  

\hline
\multirow{2}{*}{LER} & Touschek & 1.62 &  0.63  & 2.57 \\ 

 & Beam-gas &  29.39 &  2.79  & 10.53 \\ 
\hline

\end{tabular}
\caption{Comparison of combined detector data/MC ratios, excluding PLUME. Averages are calculated first by taking the mean of all channels in each BEAST or Belle~II detector, and then combining them into an average of averages.}
\label{table:avg_of_avg}
\end{center}
\end{table*}

\subsection{Luminosity Backgrounds}

After collisions began in Phase~2, attempts were made to measure the level of luminosity backgrounds generated. However, the amounts of Touschek and beam-gas backgrounds were significantly larger than the luminosity contribution and there is no conclusive observation of luminosity-induced backgrounds.

\subsection{Track and Vertex Distributions}
\label{sec:trk_vtx_distributions}
\index{Track and Vertex Distributions}
\def\belletwo {Belle~II} 

In addition to the reconstruction of physics events from collisions, the Belle~II tracking and vertexing software can be used to study decay vertices from beam backgrounds and $\gamma\to e^{+}e^{-}$ conversions in material away from the IP. Pairs of tracks identified by the inner tracker (PXD, SVD) and drift chamber (CDC) are assigned a pion mass hypothesis and vertexed with the RAVE algorithm \cite{rave}. The vertex positions were then examined in colliding and single beam runs to determine the location and intensity of beam particle losses and secondary interactions.

\begin{figure*}[htp]
\begin{center}
\includegraphics[width=0.98\textwidth]{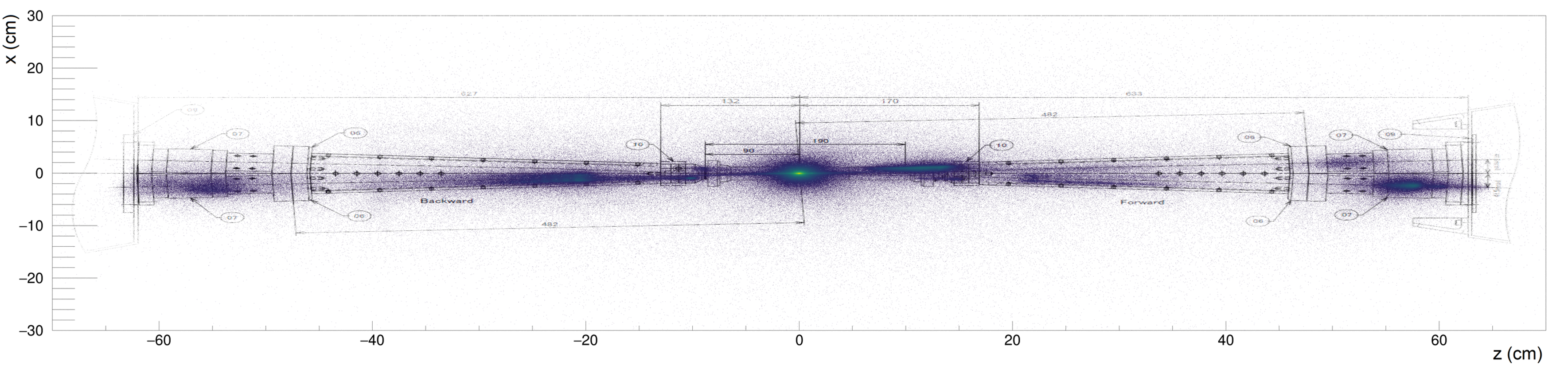}
\end{center}
\caption{(color online) Reconstructed vertices of charged tracks in Phase~2, plotted in the $xz$ plane. Clustering of events indicates the location and intensity of background beam-wall interactions.}
\label{fig:zvsx}
\end{figure*}

\begin{figure}[htp]
\begin{center}
\includegraphics[width=0.48\textwidth]{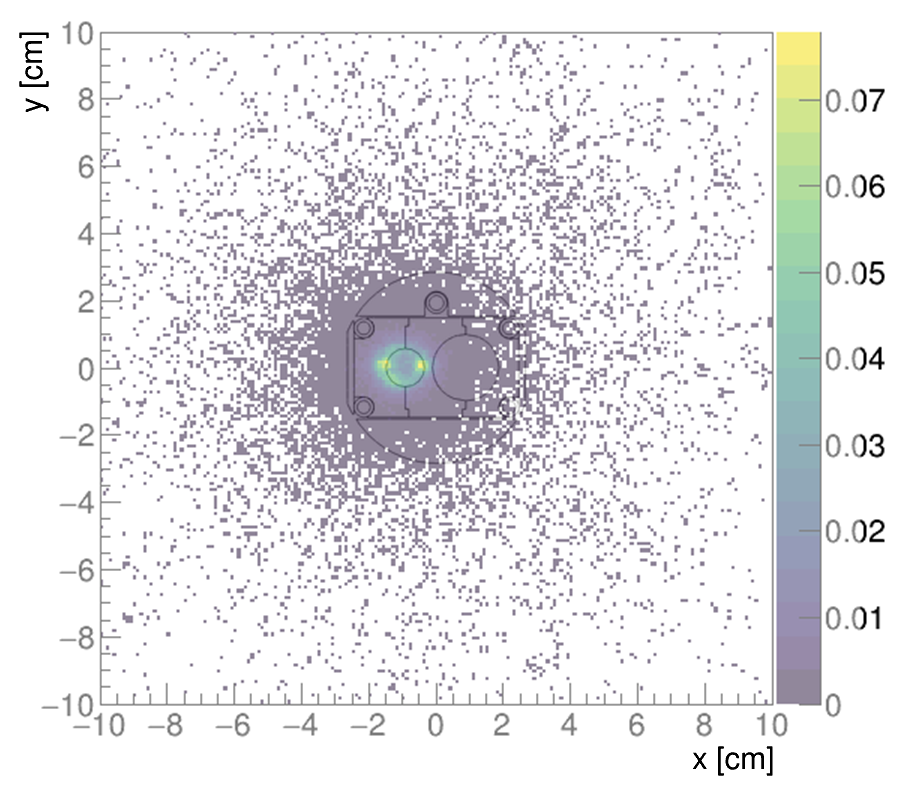}
\includegraphics[width=0.48\textwidth]{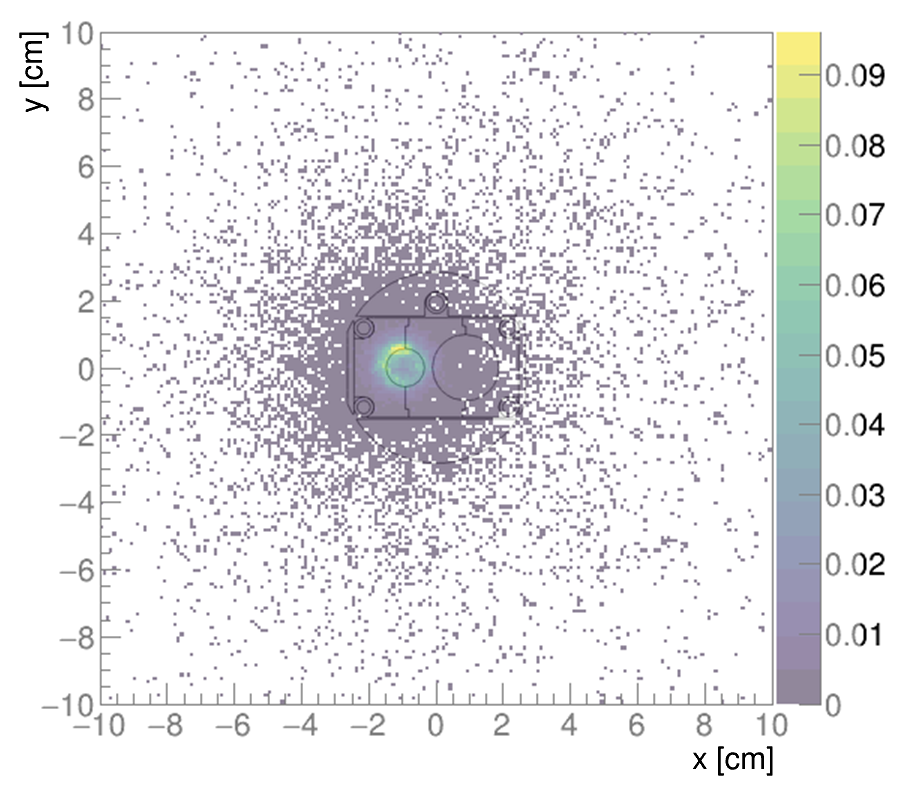}
\end{center}
\caption{(color online) Reconstructed $(x,y)$ positions of charged track vertices for HER single beam runs with differing emittance. The contour shows the envelope of the incoming HER beam. A clear change in the distribution of beam-wall interactions can be seen with the varying accelerator conditions.}
\label{fig:xvsy}
\end{figure}

Figure \ref{fig:zvsx} shows the $(x,z)$ location of reconstructed vertices for Phase~2 physics runs, where $z$ refers to the direction along the Belle~II solenoid axis. To aid in visualization, a minimum intensity threshold has been applied and a diagram of the beamline overlaid. Reconstructed vertices are concentrated at the IP as expected from $e^{+}e^{-}$ collisions, as well as at several locations along the HER and LER beam axes. Cross-sectional $(x,y)$ diagrams taken in steps along the $z$ axis illustrate the structure of these backgrounds. Figure \ref{fig:xvsy} provides a comparative example of single HER beam background data from $-30 < z < -25$\,cm recorded under two different emittance scenarios. These studies were used to detect and investigate the aforementioned ``scraping beam'' backgrounds due to non-optimal beam steering, and to provide information for mitigation strategies such as further beam tuning and future implementation of additional shielding.

\subsection{Injection Backgrounds}
\label{sec:injection_backgrounds}
This section presents time-resolved analyses of the injection-induced particle background at the IP which are the basis for the current usage of CLAWS as a beam abort system. First, we present a time resolved hit energy spectrum for each of the two main rings, illustrating the ability to detect irregularities in the beams. Subsequently, we investigate the impact of the injection bunch on neighboring bunches, demonstrating the precise timing capability of the CLAWS system.
\begin{figure}[h!]
  \centering
  \includegraphics[width=0.45\textwidth]{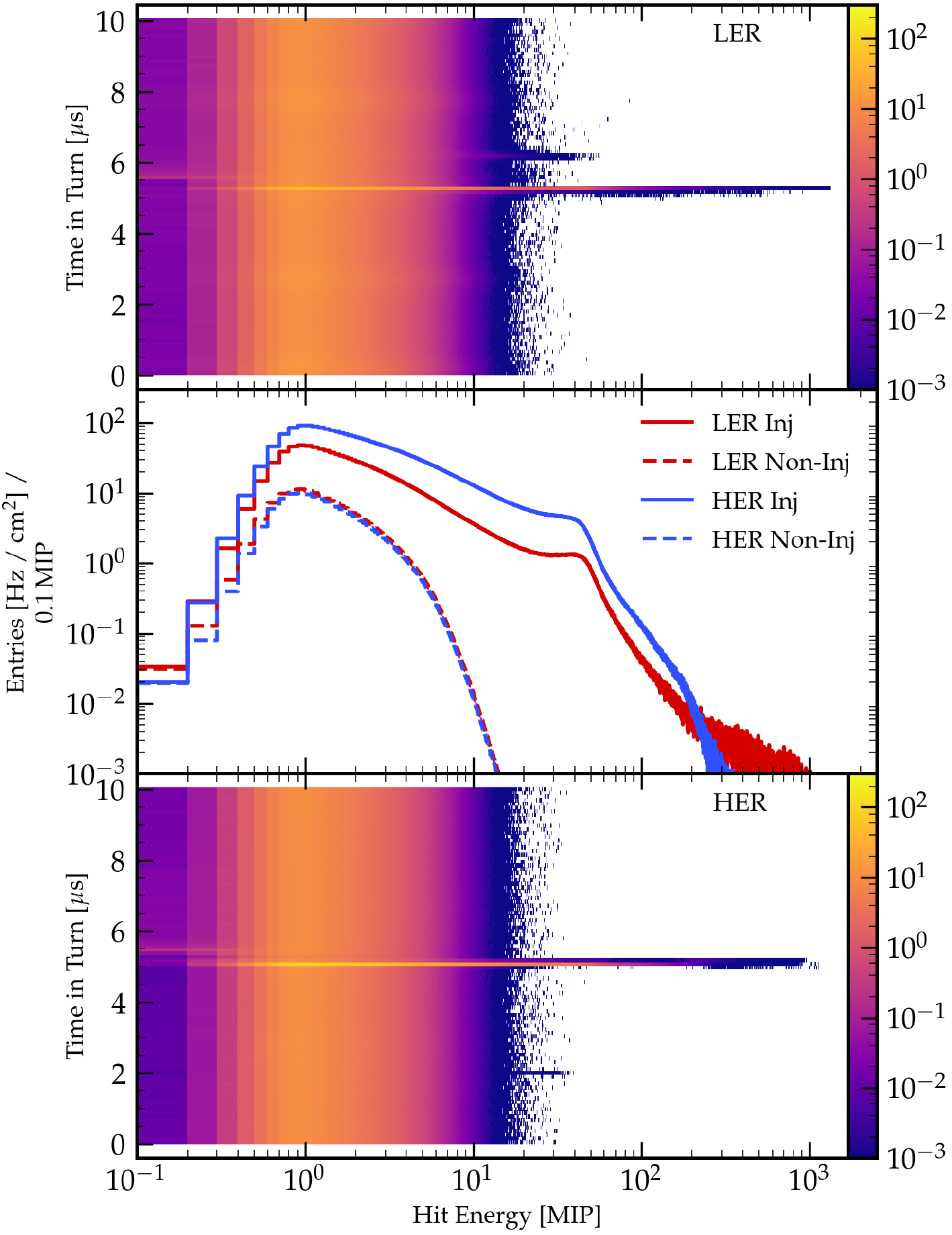}
  \caption{(color online) Hit energy spectra in the CLAWS detector. (top) and (bottom) Time resolved hit energy spectrum of the LER and HER, respectively. (middle) Projection of the top and bottom plot with marginalized time in turn variable separated by injection and non-injection times of a revolution. The units of the color bars are equal to the vertical axis of the middle plot. The plot uses all injection data collected by CLAWS between May 25 until the end of Phase 2.}
  \label{fig:claws_her_hit}
\end{figure}
\par
CLAWS signals are measured as hits with the unit of MIP, equivalent to the most probable energy deposition of a minimum-ionizing particle. The amplitude of a hit is a proxy for the number of particles crossing one CLAWS sensor simultaneously at the time of the hit. \autoref{fig:claws_her_hit} (top) and (bottom) illustrate the time-resolved hit energy distributions. The \emph{time in turn} variable displayed on the $y$ axis assigns a relative time to each circulating bunch at which it passes by the IP. This pattern repeats every revolution with $T_{rev} = \SI{10.0614}{\micro\second}$, assuming a fixed bunch fill pattern. The starting time is chosen arbitrarily. Therefore, we obtain the time in turn with \mbox{$t_{\text{turn}} = t_{\text{signal}}\,\mathrm{mod}\,T_{\text{rev}}$}. The horizontal excesses around $t_{\text{turn}} = \SI{5.5}{\micro\second}$ contain the signals of the injected bunch passing the IP. The low energy excess following the injection bunches in \autoref{fig:claws_her_hit} (top) and (bottom) are SiPM intrinsic afterpulses. All other hits emerge from stable circulating bunches. The projections of the injection and the average of all non-injection times of one revolution are illustrated in \autoref{fig:claws_her_hit} (middle).
\par
The time dependent hit energy spectra in \autoref{fig:claws_her_hit} (top) and (bottom) demonstrate the confinement of the injection induced backgrounds to a narrow time window during each turn. Highly energetic hits above $10\,\text{MIP}$ emerge almost exclusively from the newly injected bunches. Nonetheless, for each ring a second, much less pronounced, high energy excess is observable. The second excess in the HER at around $t_{\text{turn}} = \SI{2}{\micro\second}$ is the result of a single event that results in hits in multiple detectors. In contrast, the second excess in the LER around $t_{\text{turn}} = \SI{6.25}{\micro\second}$ originates from a period of approximately $5\,\text{h}$ on June 19, 2018, with signals up to $60\,\text{MIP}$ consistently observed \SI{897\pm 51}{\nano\second} after the fifth pass-by of the injection bunch. This suggests a connection of this phenomenon to injections, but it was not possible to identify its origin.
\par
\autoref{fig:claws_her_hit} (middle) demonstrates that HER \emph{(HER Inj)} injections cause a higher rate in the CLAWS sensors compared to injection bunches in the LER \emph{(LER Inj)}. For low hit energies from $1\,\text{MIP}$ to $100\,\text{MIP}$ the HER rate is around \SI{100}{\hertz\per\centi\meter\squared}, a factor of 4 higher than the respective rate from the LER, indicating larger numbers of background particles originating from LER injections.  Around $300\,\text{MIP}$ the LER rate becomes higher than the HER. 

In general, injections in the LER result in higher hit energies compared to injections into the HER. The slightly higher rates of the non-injection graphs of the LER compared to the HER result from the higher average beam currents in the LER during the time of data taking.
\par
\begin{figure}[h!]
  \centering
  \includegraphics[width=0.45\textwidth]{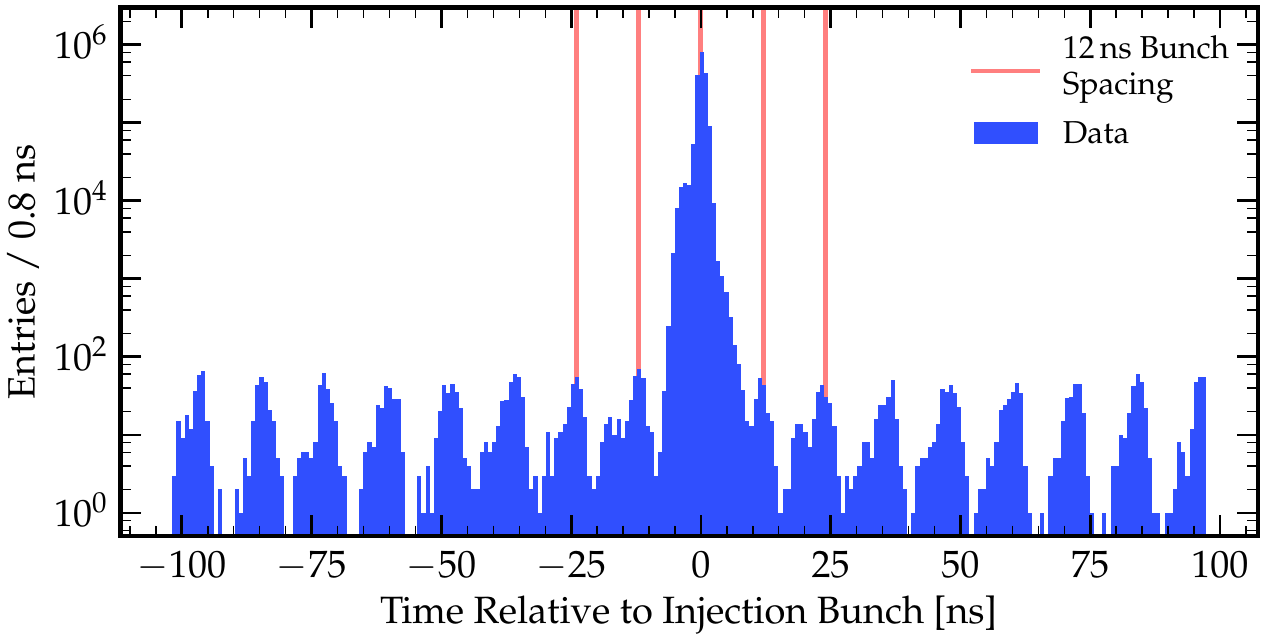}
  \caption{Arrival times of signals with a minimum energy of $2\,\mathrm{MIP}$. Each peak emerges from a circulating bunch. The red lines mark exemplarily the bunch spacing of $12\,\mathrm{ns}$ around the injection bunch. The plot uses HER injection data from July 13, 2018.}
  \label{fig:claws_her_pre_post_inj_bunches}
\end{figure}

\autoref{fig:claws_her_pre_post_inj_bunches} illustrates the time structure around the injection bunch of dedicated HER injections on July 13, 2018. Each peak originates from a circulating bunch with a peak-to-peak distance equal to the bunch spacing of \SI{12}{\nano\second}. This time-resolved analysis illustrates the increased background level during the passage of the injection bunch only. Adjacent bunches do not show any negative influence of the high background of the injection bunch.
\par
These results demonstrate the capabilities of CLAWS to observe even minimal irregularities in the circulating beams of both main rings. Together with its precise time resolution of \SI{317\pm4}{\pico\second} for hit energies of 10\,MIP, this makes the CLAWS detectors well suited to act as a beam abort system. For physics data taking,  we will implement a CLAWS-based abort system capable of issuing a fast abort signal in case of sudden increases of background, providing the signal typically one turn faster than systems relying on the measurement of integrated doses.

\par

\subsection{Synchrotron Radiation in PXD}
\label{sec:sr_results}

In addition to the backgrounds described above, there is evidence for an additional synchrotron radiation background component in the PXD. In Phase~2, the presence of SR background can be inferred from study of cluster energy spectra. In particular, photons originating from SR can be identified as low energy single-pixel clusters with a characteristic spatial distribution along the $z$-direction.   

\begin{figure}[htb!]  
\centering
\includegraphics[trim=0 0 45 0,clip,width=.9\linewidth]{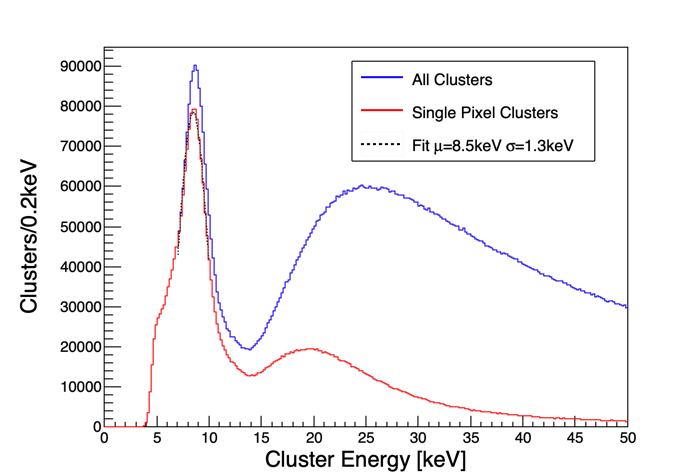} 
\includegraphics[width=0.45\textwidth]{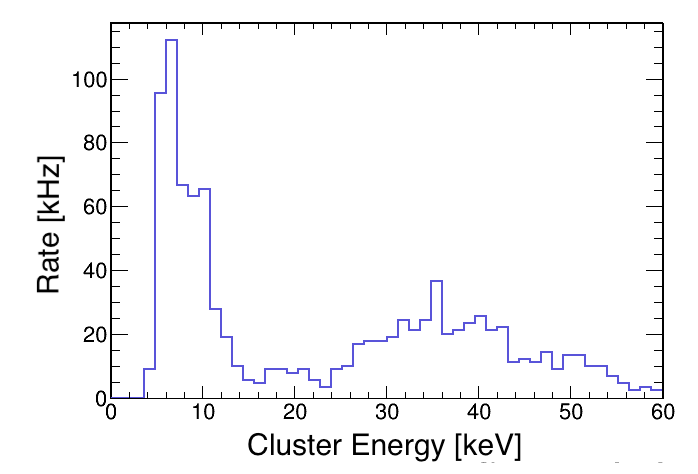} 
\caption{(color online) Measured (top) and simulated (bottom) energy spectra in the forward innermost sensor of PXD. Measured spectra are shown in blue for all clusters and in red for single pixel clusters.} 
\label{fig:if_and_of_charge_5613} 
\end{figure}

The study of cluster energy spectra in the PXD was performed with physics data using the standard PXD reconstruction algorithm. A PXD hit threshold of either 5 or 7 Analog-to-Digital converter Units was applied in the front-end electronics and the highest charge in the PXD cluster was required to be above 7\,Units. The seed charge cut of 7\,Units translates to an energy threshold of around 5\,keV. These PXD clusters are not required to be matched to a charged track from the SVD or CDC, and they can come from any charged particle or photon depositing enough energy and hitting the PXD during the 20\,\si{\micro}s time window around the trigger. The resulting calibrated energy spectra are shown separately for single and multi-pixel clusters for the PXD inner forward sensor in~\autoref{fig:if_and_of_charge_5613} (top), with the corresponding simulated energy spectrum, normalized to an HER current of 250\,mA and with the same 5\,keV energy threshold (bottom). This simulated energy spectrum can be compared with measuremed data (presented in red) only qualitatively as there are two substantial differences: firstly, this simulation only considers {\SR} photons originating from the HER beam, whereas the measured energy spectra contain contributions from both the HER and LER; Secondly, the measured energy spectra contain both photons and a non-negligible contribution from charged particles. 
The peak around 10\,keV, which is seen both in simulated and measured energy spectra, can be explained by \SR-induced fluorescence photons emitted perpendicular to the $z$ direction from the gold layer on the inner surface of the beam pipe. Such photons are capable of penetrating the beryllium beam pipe. The simulated broad distribution peaking at larger energy of 35\,keV is produced by {\SR} photons which penetrate the beam pipe directly. This component is less visible in the measured energy spectra because of the much larger additional component due to charged particles, which peaks around 20\,keV.


The energy spectra can be split into a soft component ($<$10~keV) and a hard component ($>$10~keV). The hard part of the spectrum contains more multi-pixel than single pixel clusters and shows a Landau peak as expected for clusters originating from charged particles. In contrast, the soft part of the spectrum has a narrow energy peak on top of the low energy tail of the Landau distribution. The narrow peak is almost exclusively formed by single pixel clusters.

 A Gaussian fit to the single pixel charge, restricted to the 7-10\,keV range, shows that the peak positions range between 8.4-8.6~keV and are nearly identical for all four PXD sensors. This last observation hints at a common origin for the narrow energy peak. The number of single pixel clusters in the energy range 7-10~keV, after subtracting a flat background, is found to be ten times higher in the first PXD layer compared to the second. This supports the hypothesis that the common origin of the soft-photon background is SR, in which the scattered photons are expected to be mostly absorbed in the first PXD layer.

\begin{figure}[htb!]  
\centering
\includegraphics[width=.9\linewidth]{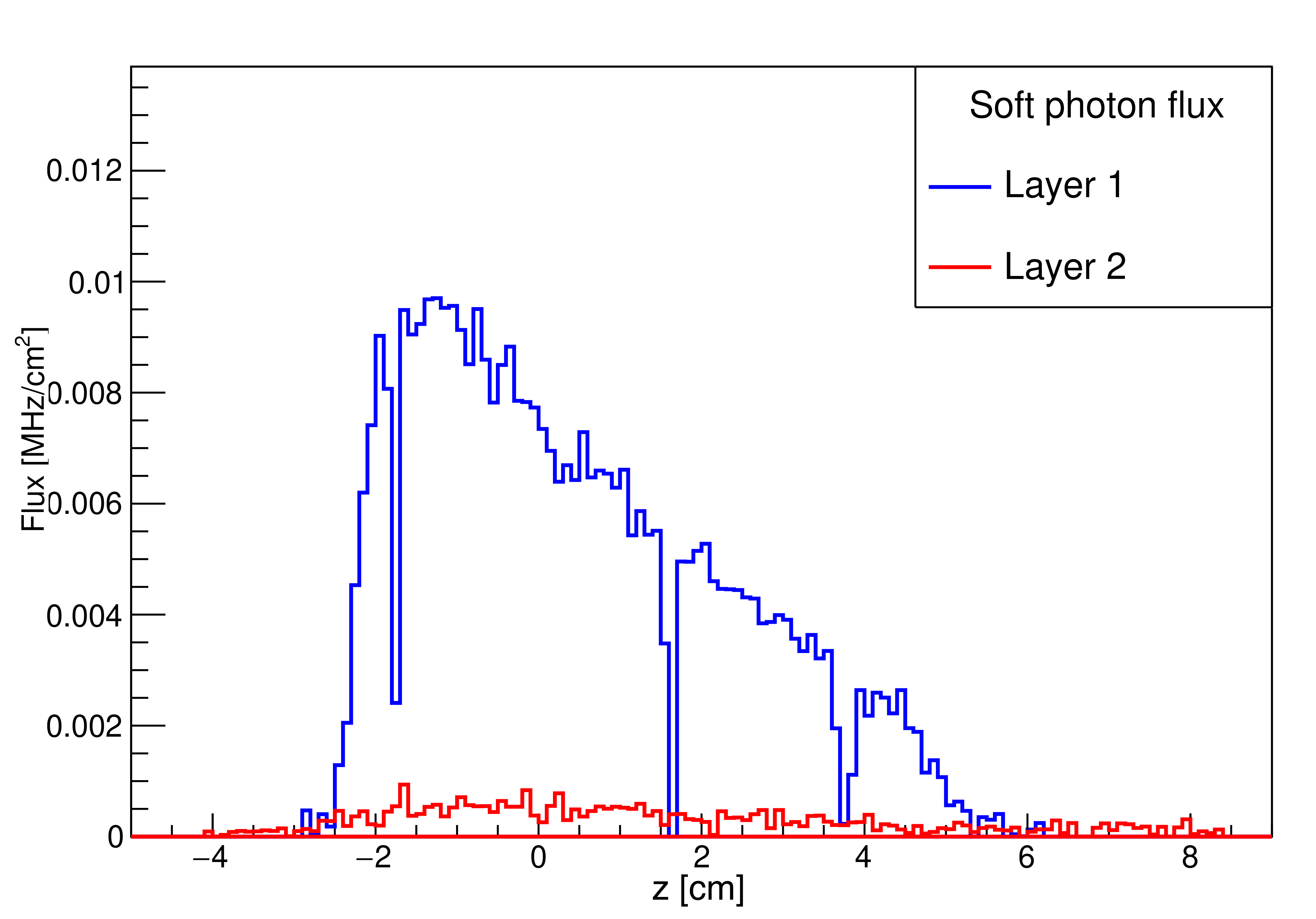} 
\includegraphics[width=0.45\textwidth]{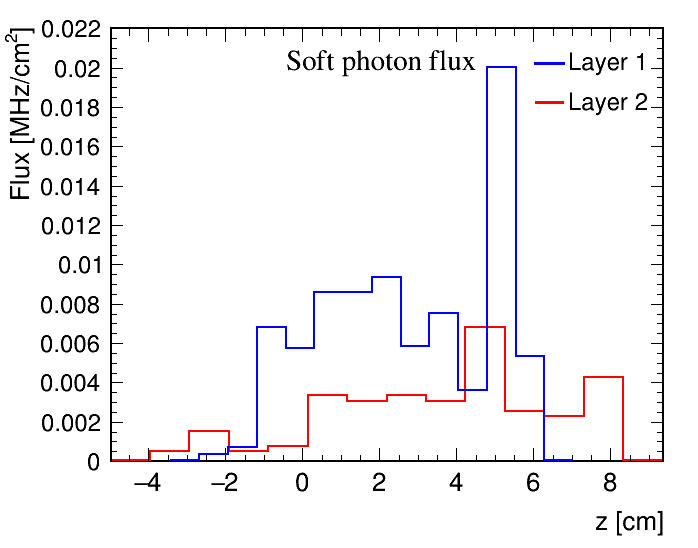} 
\caption{(color online) Measured (top) and simulated (bottom) soft photon flux on first PXD layer (Layer 1) and second PXD layer (Layer 2) against $z$-position for single-beam runs in the HER. The average HER current in the single beam run is 250\,mA, and simulation uses the same value.} 
\label{fig:soft_flux_ler_and_her} 
\end{figure}

In addition, the flux of particles onto the PXD is measured as a function of the $z$ position and layer number during dedicated single-beam runs. The spatial profile of the soft photon flux,  identified as single pixel clusters with energy $<10$ \,keV, is shown for a LER (HER) single-beam run in the top (bottom) of~\autoref{fig:soft_flux_ler_and_her}.

As seen in Figure~\ref{fig:soft_flux_ler_and_her}, The simulated (top) and measured (bottom) spatial profiles are found to be incompatible. Dedicated simulations of low-energy SR photons indicate the presence of an excess of events caused by photons forward scattering off the ingoing copper-coated Tantalum beam pipe and subsequently inducing fluoresence in the gold layer on the central beam pipe. For Phase~3, the impact of this enhanced low-energy component will be significantly reduced by increasing the thickness of the gold layer from 6.6 to 10\,\si{\micro}m. 

\subsection{PXD-Based Total Ionizing Dose Estimate}
\label{subsec:pxd_dose_estimation}
We use the PXD system to estimate the total ionizing dose rate in the VXD during Phase~2 operation. Using simulations as reference, we perform an energy calibration of the measured PXD hits and clusters. Using this calibration we calculate the energy deposition inside the PXD, and then the dose per cluster.
\par
Taking into account the 20\,$\si{\micro}$s readout time of the PXD sensors, we calculate a dose rate with 1\,s binning and add it to the aforementioned common output files of the BEAST detectors. This technique assumes that the difference between data taken with and without selecting for physics events is negligible, which is justified as the luminosity during Phase~2 was significantly lower than is expected in the final experiment. 
\par
The calculated PXD dose rate was used as a second source for the dose rate in addition to the VXD radiation-monitor and beam-abort system (see Section~\ref{subsubsec:diamond_system}) that primarily provided information about the dose rate during Phase~2 operation. 
\par
Analyses show that the ratio of the PXD and diamond dose rates is not constant but depends on the accelerator beam current. While the PXD/diamond ratio was not constant over time, we find a heuristic relation between the two measurements incorporating the beam currents.

\par

In contrast to the diamond system, the PXD did not take data continuously during Phase~2 operation and only covered $\approx~12$\% of the total operation time. To fill in the gaps in the data set, we use the diamond data as well as the beam currents to infer a PXD dose rate based on the heuristic relation between the PXD and diamond data set. A more detailed description of this analysis can be found in Ref.~\cite{SchreeckPhD:2020}.
\par
\begin{figure}[htb]
\centering
\includegraphics[width=0.5\textwidth]{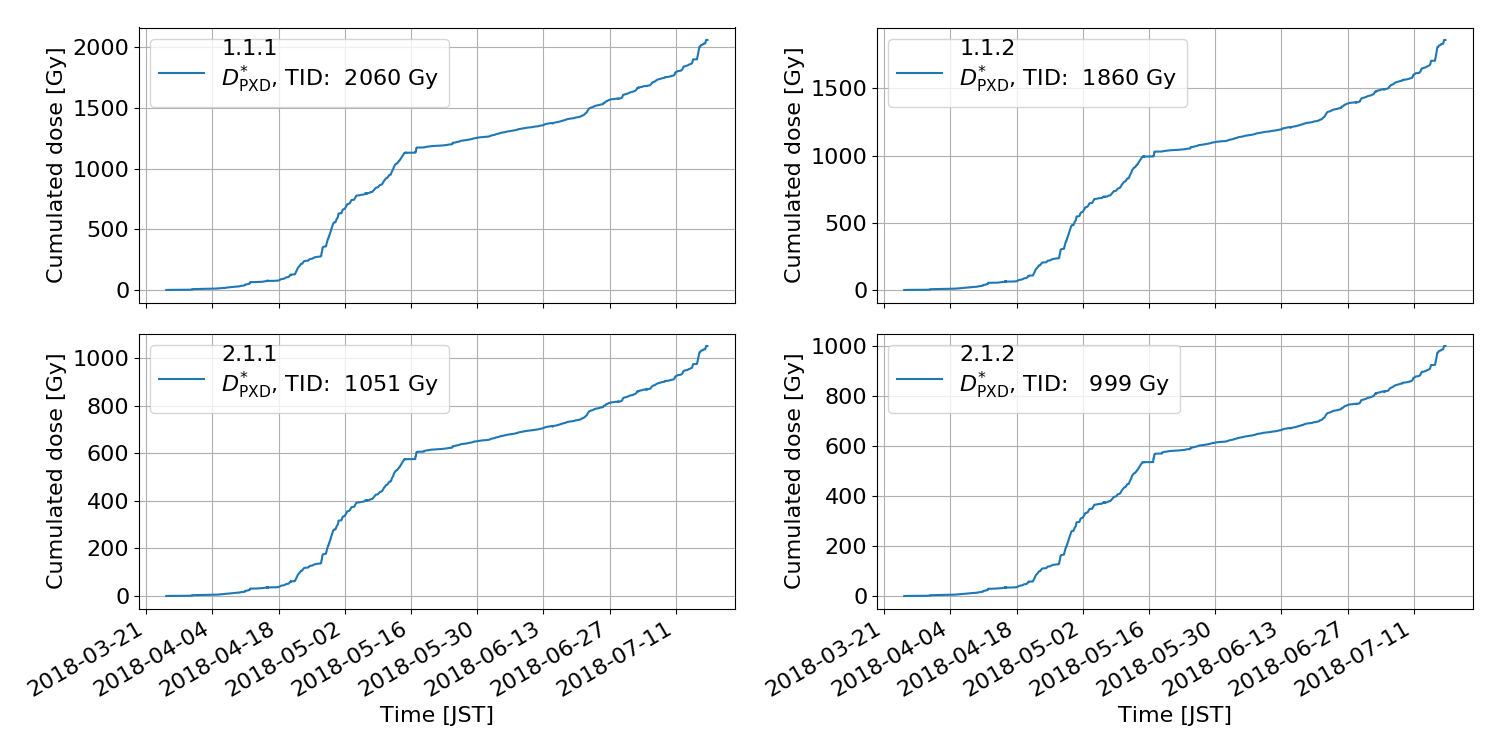}
\caption{Cumulative PXD radiation dose versus time for the four individual PXD sensors. The inner sensors (top) received approximately twice the dose compared with the outer sensors (bottom).}
\label{fig:pxd_accum_dose}
\end{figure}

\begin{figure}[htb]
\centering
\includegraphics[width=0.5\textwidth]{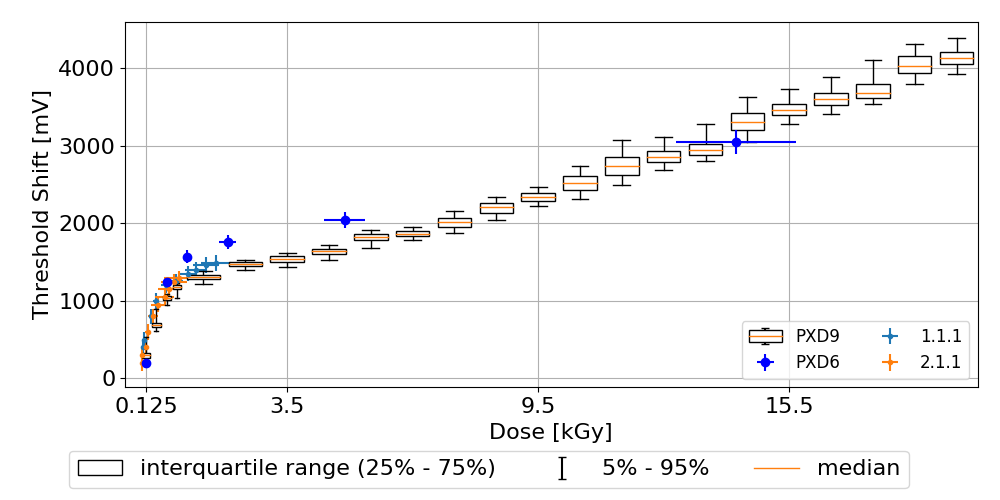}
\caption{PXD threshold voltage shifts versus cumulative dose.  Two of the PXD Phase~2 sensors (1.1.1 and 2.1.1) are compared with results gathered during two irradiation campaigns (PXD6 and PXD9~\cite{RITTER201379}).}
\label{fig:pxd_threshold_voltage_shift}
\end{figure}

 
Using a combination of the measured and calculated PXD dose rates we determine a total dose for all four PXD sensors as a function of time, which can be seen in Figure~\ref{fig:pxd_accum_dose}.
The DEPFETs (DEpleted P-channel Field Effect Transistors) of the PXD sensors show a threshold voltage shift when exposed to ionizing irradiation. 
Figure~\ref{fig:pxd_threshold_voltage_shift} shows the measured threshold shift of two of the four PXD sensors as a function of the calculated PXD dose. In addition, the threshold shift curves recorded at two irradiation campaigns, one with a prototype \cite{RITTER201379} and one with final PXD sensors \cite{Schreeck2020}, are shown. The comparison shows good agreement between the curves.
\par
Based on the calculation described above, the inner PXD sensors received a dose of $\approx2$\,kGy, and the outer sensors $\approx1$\,kGy, between March and July 2018. We deem these levels to be acceptable for the installation of the VXD for Phase~3.



\section{Collimator Study}
\label{sec:collimator_study}
\index{Collimator Study}

\begin{figure}[h]
    \begin{center}
        \includegraphics[width=.4\textwidth]{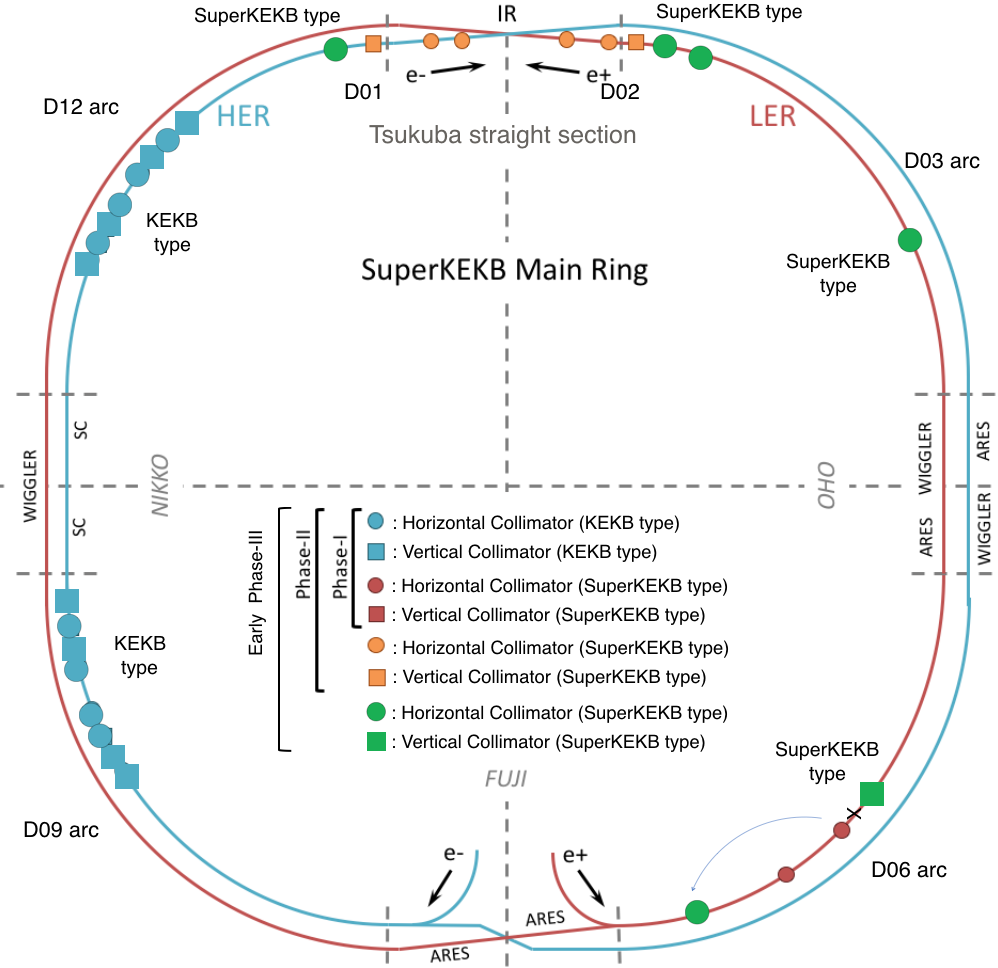}
        \caption{Map of collimators in SuperKEKB. Collimators introduced before Phase~2 are shown in orange, and those installed between Phase~2 and Phase~3 are shown in green.}
        \label{fig:coll_early_phase3}
    \end{center}
\end{figure}

SuperKEKB has a number of horizontal and vertical collimators distributed along the beamline as shown in Figure~\ref{fig:coll_early_phase3}.
As noted in Section \ref{section_introduction}, Touschek and beam-gas backgrounds are sensitive to collimation, and thus adjustment of collimator apertures can significantly reduce backgrounds.

During beam operation, collimators are widened or narrowed to remove the off-orbit beam particles away from the beam core, particularly in the IR. However, even collimators far upstream of the IP may have a significant effect on beam behavior and beam-induced backgrounds in Belle~II. To account for these effects and simultaneously determine the optimal collimator configuration during operation, we performed a number of dedicated collimator studies.

SuperKEKB collimators are implemented in the SAD simulation. In order to optimize the collimator aperture and to have a comparison between operations and simulation, a collimator simulation study was carried out prior to Phase~2 operation.

\subsection{Collimator Optimization Studies During Operations}

In keeping with the objective of determining the optimal opening width (aperture) for each collimator along the SuperKEKB beamline, collimators were systematically opened or closed in succession, after which the beam backgrounds and lifetimes were observed to determine the effect of the change.

These studies were carried out with single beams (i.e. HER or LER only), using the following procedure:

\begin{enumerate}
\item Collimators had an initial (non-optimized) setting determined during initial operations in order to protect the Belle~II detector from beam backgrounds.
\item Immediately before collimator studies, all collimators of the studied beam were slightly opened from their nominal aperture setting, so that the physical aperture was slightly larger than the narrowest point in the beam line, corresponding to the QCS components.
\item Starting from the collimator closest to the beam injection system and going toward the interaction point (IP), each collimator was closed in fixed steps. After several minutes, a top up injection was performed in order to verify the effect on injection background. After injection was completed, storage background was observed for a few minutes. This procedure was iterated for the same collimator until beam lifetime worsened with no significant improvement on the background.
\item The same procedure was applied to all collimators, until an initial optimal setting was found. 
\end{enumerate}

A second goal of the study, together with the reduction of beam background in the interaction region, was to compare the optimal collimator setting found during collimator studies with that obtained in simulation. The comparison gives important indications of the accuracy of the simulation and possible improvements.

\subsection{SAD Simulation Studies}

The collimator study using SAD was carried out with the following procedure:

\begin{enumerate}
\item The starting point is the simulation with all collimators of the lattice fully open.
\item Each collimator is individually closed in steps. For each step a short version of the simulation (to save storage and computing resources) is run, to see the impact of the new collimator width on IR background and beam lifetime. The optimal width for each collimator is found when the increase in the total ring loss rate is greater than the decrease in the IR loss rate.
\item A full simulation is run with all collimators set at the optimal aperture found in the previous step, to evaluate the full impact of all collimators closed together at their optimal aperture.
\item The last collimator setting obtained is used as the starting point for a second round of the same procedure. Each collimator is individually closed further, and a short version of the simulation is used to evaluate the impact of this modification.
\item The optimization procedure concludes if no further improvement on the IR background is possible without affecting beam lifetime. If the IR background is still higher than the limit of 100\,MHz, the collimator settings are modified to reduce the IR background, even if this affects beam lifetime.
\end{enumerate}

\subsection{MC Optimization for Early Phase~3 and Comparison Against Phase~2}
\label{sec:mc_op}
For operation in early Phase~3, more collimators will be added to reduce beam backgrounds further. We perform another collimator study using SAD with accelerator parameters and collimators planned for early Phase~3. In order to directly compare simulations of Phase~2 and early Phase~3, which involve different beam parameters, the Phase~2 simulation results were scaled to the early Phase~3 parameters using Eq.~1.




As shown in Table \ref{tab:losses_table}, a comparison of simulations shows that if the optimal collimator setting found in simulation can be reproduced during early Phase~3 operations, with the installation of new collimators, the following IR background reduction factors can be achieved: LER backgrounds can be reduced by a factor of 4.4, from 346\,MHz to 78.2\,MHz, and HER backgrounds can be reduced by a factor of 3.7, from 34\,MHz to 9.3\,MHz.


\begin{table*}
\begin{center}
\begin{tabular} {| c | c | c | c | c |}
	\hline
	& \multicolumn{2}{|c|}{Late Phase~2} & \multicolumn{2}{|c|}{Early Phase~3}\\
	\cline{2-5}
	& LER & HER & LER & HER \\
	\hline
	IR losses - Coulomb (MHz) & 186 & 1.2 & 28.6 & 0.4\\
	IR losses - Touschek (MHz) & 160 & 32.8 & 49.6 & 8.9\\
	\hline
	Ring losses - Coulomb (MHz) & 6116 & 1532 & 8211 & 2022\\
	Ring losses - Touschek (MHz) & 75944 & 14568 & 83351 & 15127\\
	\hline
	Lifetime (s) & 1018 & 3589 & 815 & 3564\\
	\hline
\end{tabular}
\end{center}
\caption{  Simulated beam particle loss rates for Touschek and Coulomb background components in SuperKEKB Phase~2 and Phase~3.}
\label{tab:losses_table}
\end{table*}

\section{Extrapolations and  Recommendations}
\label{sec:summary_recommendations}
\index{Summary and Recommendations}

During the course of Phase~2, we observe total beam background levels larger than simulation by factors from 1 - 100, as presented in Section~\ref{sec:ratio_summary}. By refining the simulated description of the IR geometry, we have improved the agreement between simulation and measurement, as seen by comparing the top and bottom panels of Figures~\ref{fig:tous_summary} and~\ref{fig:bg_summary}.

The remaining discrepancy in beam-gas data and simulation is unsurprising, as detailed gas composition and pressure distributions are not accounted for in the simulation. In contrast, we expect better agreement for Touschek backgrounds. Instead, we observe a discrepancy of up to a factor of 100, particlularly in the HER, consistent with the results from Phase~1. After the conclusion of this work, we found that this was the result of the physical collimator model used in the beam particle tracking code, which had been accurate for the reused KEKB collimators but did not fully describe the newly installed SuperKEKB-type collimators. This has been resolved for future studies and a detailed description is available in Reference~\cite{natochii2021improved}.

We also observed unexpected backgrounds attributed to beam scraping on beam pipe material, which we reduced largely by beam-orbit tuning and collimator optimization.

Luminosity backgrounds provided a negligible contribution to the overall background level, given the low luminosities recorded during commissioning and the high levels of other backgrounds. 

While ratios of data to simulation are useful in assessing simulation accuracy, we also utilize them to correct the simulation for the purposes of extrapolating backgrounds. By doing so, predictions remain robust even with large discrepancies in agreement, as observed in Phase~2. 


\subsection{Extrapolation to Early Phase~3}
\label{sec:extrapolation}
\index{Extrapolation to Early Phase~3}

During the course of Belle~II physics data taking, SuperKEKB plans to both increase beam currents and continue squeezing the beams at the IP, leading to a progressively more challenging background environment. Because we expect conditions at the target luminosity to be significantly more difficult than those in the early stages of data taking, it is most reasonable to mainly consider conditions in the first years of Phase~3, before conditions diverge significantly.

The most pressing question to answer with Phase~2 data is the safety of the sensitive VXD. We use measurements taken during Phase~2 to predict the expected level of background incident on the area around the IP, thus providing a meaningful measure of the risk for the PXD and SVD. In the event that measured and predicted levels are too high, it would be unsafe to install the VXD and physics data taking with the full set of Belle~II subdetectors would not be possible. 

\begin{figure*}[htb]
\begin{center}
\includegraphics[width=0.423\textwidth]{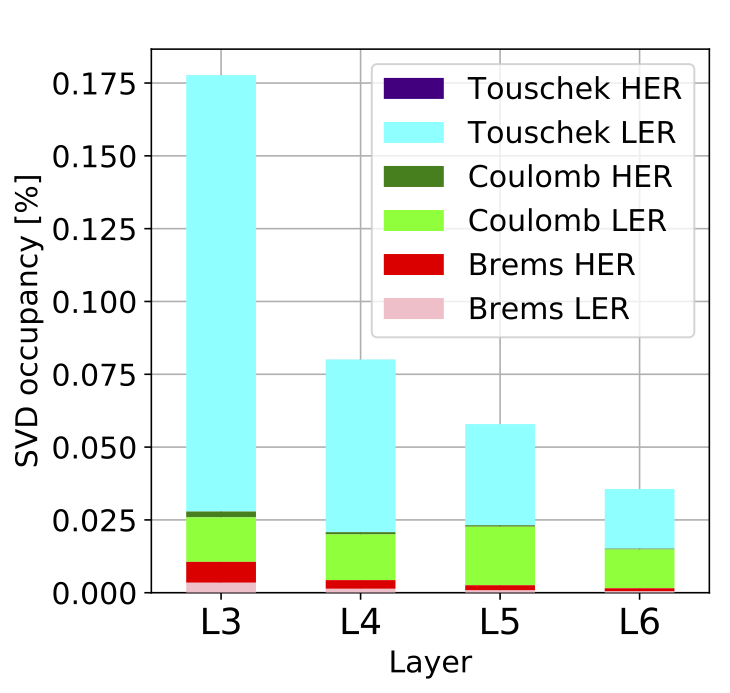}\,
\includegraphics[width=.4\textwidth]{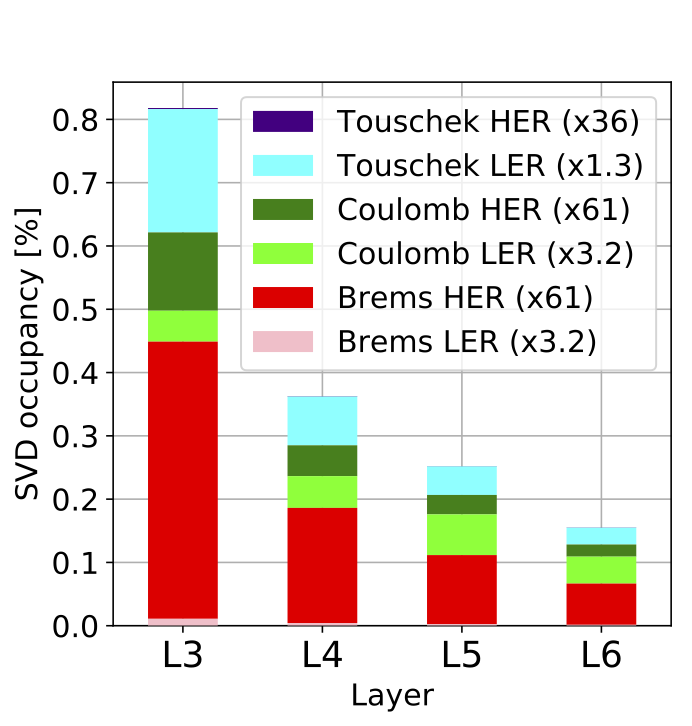}

\caption{(color online) Predicted SVD layer occupancy in early Phase~3. Left: Raw Phase~3 simulation. Right: Early Phase~3 simulation rescaled by Phase~2 data/MC ratios.}
\label{fig:svd_unscaled}
\end{center}
\end{figure*}

\begin{figure*}[h!]
   \begin{center}
     \includegraphics[height=.45\textwidth]{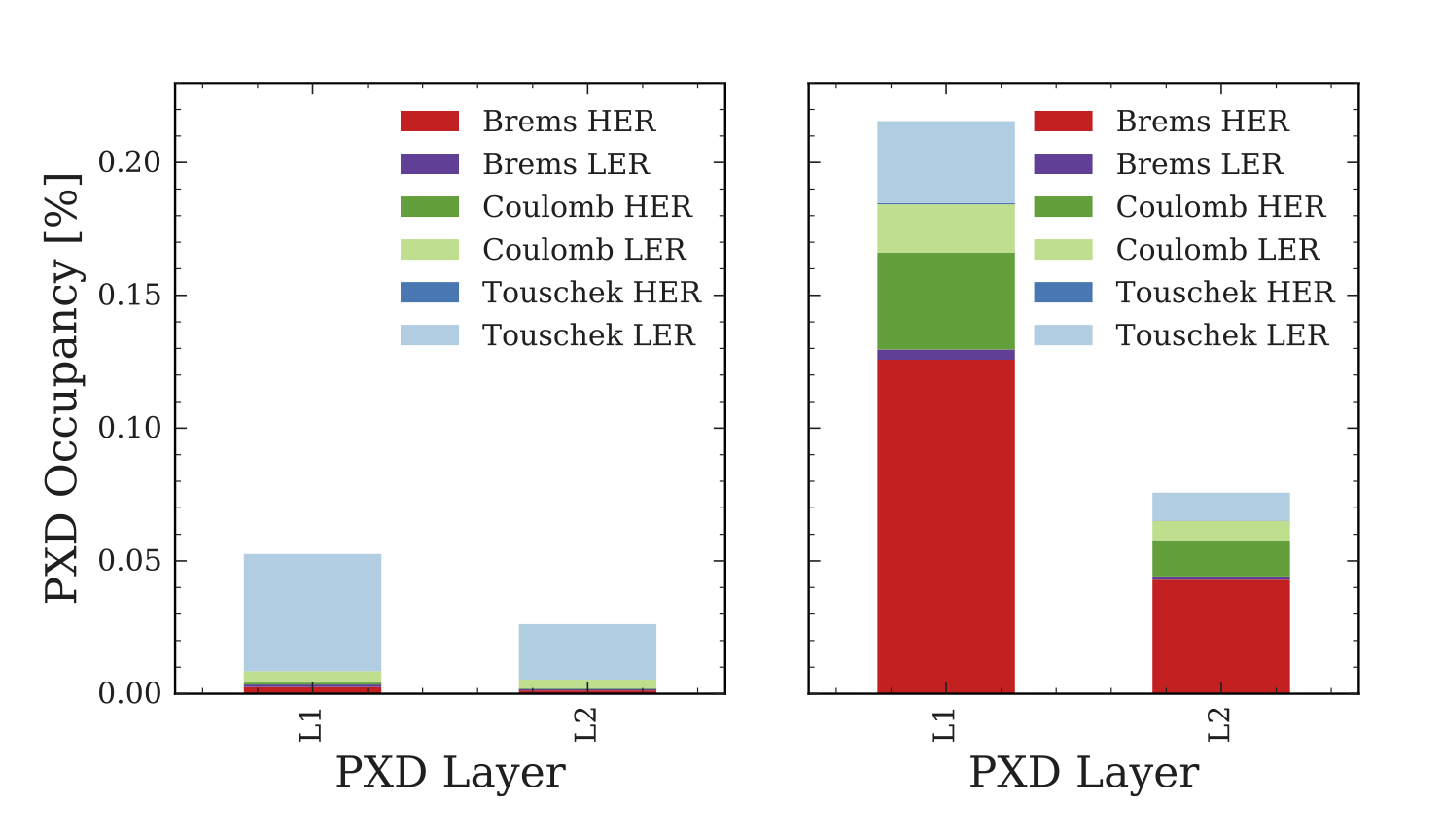}
     
     \caption{(color online) Predicted PXD layer occupancy in early Phase~3. Left: Raw Phase~3 simulation. Right: Early Phase~3 simulation rescaled by Phase~2 data/MC ratios.}
\label{fig:pxd_scaled}
\end{center}
\end{figure*}

The reduction of LER backgrounds described in Section~\ref{sec:mc_op} is critical for keeping background rates at safe levels. Figure~\ref{fig:svd_unscaled} shows the predicted occupancy levels in the SVD, extrapolated to early Phase~3 and adjusted for updated beam optics. Absent optimization of collimators and the introduction of a vertical collimator in the upstream D06 section of the LER, these rates would likely exceed the safe operating limit of the SVD. With the recommended changes, however, we predict that occupancies will remain at the level of approximately 1\%. Likewise, predicted PXD occupancies are expected to remain under 1\%, as shown in Figure~\ref{fig:pxd_scaled}. The performance of these detectors, and of trackfinding, suffers with occupancies above 3\%. Hence, the overall predicted background rates were determined to be safe.

\begin{figure}[h]
\begin{center}
\includegraphics[trim=05 0 55 45, clip,width=0.45\textwidth]{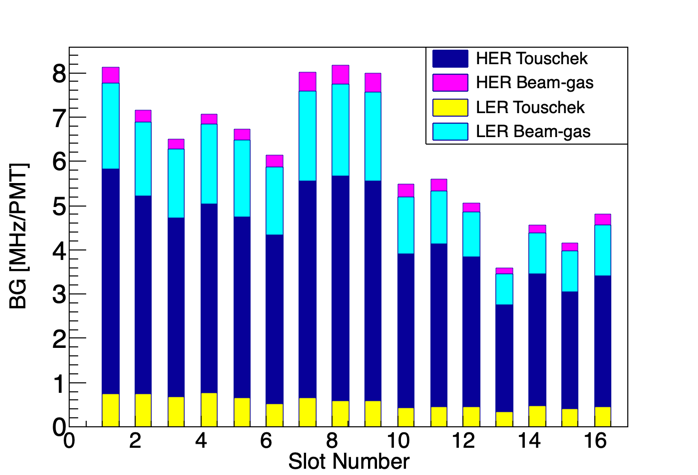}
\caption{Predicted TOP detector photomultiplier hit rates for beam-gas and Touschek backgrounds, versus TOP slot position, at full design luminosity. Belle~II TOP slots are distributed azimuthally (in $\phi$) around the SuperKEKB beam axis. The goal is to reduce the sum of these background components further, to about 2.0 MHz per PMT. See text for further discussion.}
\label{fig:top_phi}
\end{center}
\end{figure}

\subsection{Extrapolation to Design Luminosity}

Finally, we extrapolate backgrounds to the full SuperKEKB design luminosity of  $8 \times 10^{35}\,\mathrm{cm}^{-2}\,\mathrm{s}^{-1}$. This is a more challenging scenario than the target luminosity of \SI{6.5e35}{cm^{-2}s^{-1}} discussed earlier, but we make the choice because simulations and collimation strategies are more mature for the design scenario. We first simulate the machine at design beam currents and optics, using a simulation-optimized collimator setup specific to those optics. Then, we conservatively re-scale the Belle~II rates of each simulated single-beam background component by the corresponding data/MC factors measured in Phase 2. As an example, we show the predicted background composition for the TOP detector in Figure~\ref{fig:top_phi}. TOP is considered particularly vulnerable to beam backgrounds due to the finite lifetime of the photomultipliers (PMTs) used, which degrade as photocathode charge is accumulated. It is therefore critical to monitor the TOP background composition and use targeted mitigation strategies such as collimation to minimize TOP backgrounds and thereby maximally extend PMT lifetimes. At design luminosity, TOP PMT rates are expected to reach order 8\,MHz per PMT from beam collisions alone; because these events arise from beam interactions at the IP, they are affected by neither collimation nor vacuum scrubbing and hence are to first order irreducible. Background events from the single beam processes shown in Figure~\ref{fig:top_phi} contribute further to the overall PMT rates.

The Belle~II goal is to further reduce these single-beam rates to a total of order 2\,MHz by utilizing vacuum scrubbing, shielding, and collimation. While HER Touschek scattering appears to be the most critical component based on Phase 2 measurements, as will be discussed below, recent improvements in the simulation of SuperKEKB-type collimators have reduced the systematic error on the magnitude of the Touschek background forecast. The dominant background and main mitigation target at the time of writing is therefore the second-largest component in Figure~\ref{fig:top_phi}, LER beam-gas. Reducing that background is one of the main SuperKEKB challenges on the road towards design luminosity in the coming years.

Recently, the pressure in the LER has reduced substantially due to vacuum scrubbing from baseline machine operation. As a result, the LER beam-gas background may be as much as one order of magnitude lower than that shown in Figure~\ref{fig:top_phi} by the time SuperKEKB reaches its target luminosity.

\subsection{Summary and Recommendations}

Ultimately, based on the background measurements in Phase~2 and extrapolations forward to Phase~3 conditions, we made the recommendation to proceed as planned with the installation of the VXD and start of physics data taking runs in Phase~3. 

Furthermore, we also recommend that a new collimator be installed in SuperKEKB before the start of Phase~3 to reduce the dominant LER beam-gas and Touschek backgrounds. Based on the studies described in Section~\ref{sec:collimator_study}, we determine that the most beneficial place for a new LER vertical collimator is in Section D06, near the MR injector. Following completion of Phase~2, both recommendations were implemented during the shutdown period between phases.

\section{Conclusions}
\label{sec:Conclusions}
\index{Conclusions}

The second commissioning phase of SuperKEKB saw major advances in accelerator operation, including beam luminosities of $5.6 \times 10^{33}$\,cm$^{-1}$\,s$^{-1}$ and first beam collisions, and corresponding increases in beam-induced backgrounds. To meet the demands of this more challenging environment, we successfully employed both Belle~II and dedicated BEAST II detectors to measure and mitigate backgrounds due to Touschek, beam-gas, injection, and synchrotron radiation.  We carried out detailed measurements of neutron fluxes and dose rates. Finally, we performed searches for, but did not conclusively detect, luminosity backgrounds.

BEAST II successfully demonstrated numerous new technologies: FANGS, PLUME, and CLAWS detectors for measurements of incident particles and doses near the IP, directional neutron recoil monitors, and a diamond-based monitoring system that was developed to carefully track radiation doses and trigger beam aborts in the event of unacceptably high rates.  Many of these systems are now used routinely in Phase~3.

The diamond sensor beam-abort system in particular has been improved to provide faster response times to high background levels, with criteria updated to include ``fast'' and ``slow'' thresholds. These improvements have proven to be effective in preventing QCS quenches and subsequent high doses of radiation on the sensitive vertex detectors during Phase~3 data taking.

We updated the Geant4 model of the Belle~II interaction area to more closely match reality, and as a result the ratio of observed to predicted single-beam background rates improved by factors of 1.12 - 14.98, indicating improved predictive ability for future conditions. Extrapolations of measurements to Phase~3 running conditions predict background levels to be within acceptable limits provided beam collimation is improved. Based on these results, we recommended proceeding with the installation of the sensitive vertex detectors along with an additional LER collimator.

As Belle~II continues in its data collection phase and the background conditions become increasingly challenging, informative studies such as those described here will continue to enable targeted mitigation for the duration of the experiment.

\section{Acknowledgments}
\label{sec:Ack}
We acknowledge support from the U.S. Department of Energy (DOE) via
Award Numbers DE-SC0007852, DE-SC0010504, DE-AC02-05CH11231, via the
U.S. Belle~II Project administered by Pacific Northwest National
Laboratory (DE-AC05-76RL01830), and via U.S. Belle~II Operations
administered by Pacific Northwest National Laboratory and Brookhaven
National Laboratory (DE-SC0012704). Pacific Northwest National Laboratory is managed and operated by the Battelle Memorial Institute. We acknowledge the financial support by the Federal Ministry of Education and Research and by the DFG Excellence Cluster “Origin and Structure of the Universe” of Germany. This work was partially supported by the European Union’s Horizon 2020 Research and Innovation programme under Grant Agreement no. 654168 and the Horizon 2020 Marie Sklodowska-Curi RISE project JENNIFER grant agreement No. 644294. The Strasbourg Group acknowledges support from L'Institut National de Physique Nucl\'eaire et de Physique des Particules (IN2P3) du CNRS (France) and Investissements d’Avenir and Université de Strasbourg (IdEx grants W15RPE12 and W17RPD30). We acknowledge the support of Grant CIDEGENT/2018/020 of Generalitat Valenciana (Spain).

\section*{References}

\bibliography{biblio}

\begin{thebibliography}{10}
\expandafter\ifx\csname url\endcsname\relax
  \def\url#1{\texttt{#1}}\fi
\expandafter\ifx\csname urlprefix\endcsname\relax\def\urlprefix{URL }\fi
\expandafter\ifx\csname href\endcsname\relax
  \def\href#1#2{#2} \def\path#1{#1}\fi

\bibitem{PTEP:Ohnish}
Y.~Ohnishi, et~al., {Accelerator design at SuperKEKB}, Progress of Theoretical
  and Experimental Physics 2013 (2013) 03A011.
\newblock \href {https://doi.org/10.1093/ptep/pts083}
  {\path{doi:10.1093/ptep/pts083}}.

\bibitem{Abe:2010gxa}
T.~Abe, et~al., \href{http://cds.cern.ch/record/1304162}{{Belle II Technical
  Design Report}}, Tech. Rep. KEK REPORT 2010-1, KEK, edited by: Z. Dole\v{z}al
  and S. Uno (2010).
\newline\urlprefix\url{http://cds.cern.ch/record/1304162}

\bibitem{KEK:1995sta}
N.~Toge, et~al., \href{http://cds.cern.ch/record/475260}{{KEKB B-Factory Design
  Report}}, Tech. Rep. KEK-REPORT-95-7, KEK, Tsukuba (1995).
\newline\urlprefix\url{http://cds.cern.ch/record/475260}

\bibitem{Bona:2007qt}
M.~Bona, et~al., {SuperB: A High-Luminosity Asymmetric $e^+e^-$ Super Flavor
  Factory. Conceptual Design Report}, Tech. Rep. SLAC-R-856, INFN-AE-07-02,
  LAL-07-15, INFN-AE-07-2, INFN (2007).
\newblock \href {http://arxiv.org/abs/0709.0451v2} {\path{arXiv:0709.0451v2}}.

\bibitem{Lewis:2018ayu}
P.~M. Lewis, et~al., {First Measurements of Beam Backgrounds at SuperKEKB},
  Nucl. Instrum. Meth. A914 (2019) 69--144.
\newblock \href {http://arxiv.org/abs/1802.01366} {\path{arXiv:1802.01366}},
  \href {https://doi.org/10.1016/j.nima.2018.05.071}
  {\path{doi:10.1016/j.nima.2018.05.071}}.

\bibitem{Mulyani:IBIC2015-TUPB025}
E.~Mulyani, J.~Flanagan,
  \href{http://accelconf.web.cern.ch/AccelConf/IBIC2015/papers/tupb025.pdf}{{D}esign
  of {C}oded {A}perture {O}ptical {E}lements for {S}uper{KEKB} {X-}ray {B}eam
  {S}ize {M}onitors}, in: Proc. International Beam Instrumentation Conference,
  Melbourne, Australia, 2015, pp. 377--380.
\newblock \href {https://doi.org/10.18429/JACoW-IBIC2015-TUPB025}
  {\path{doi:10.18429/JACoW-IBIC2015-TUPB025}}.
\newline\urlprefix\url{http://accelconf.web.cern.ch/AccelConf/IBIC2015/papers/tupb025.pdf}

\bibitem{IBIC2016}
E.~Mulyani, J.~Flanagan,
  \href{http://jacow.org/ibic2016/papers/tupg72.pdf}{{C}alibration of {X-}ray
  {M}onitor during the {P}hase {I} of {S}uper{KEKB} commissioning}, in: Proc.
  International Beam Instrumentation Conference (IBIC'16), Barcelona, Spain,
  2016, pp. 525--528.
\newblock \href {https://doi.org/10.18429/JACoW-IBIC2016-TUPG72}
  {\path{doi:10.18429/JACoW-IBIC2016-TUPG72}}.
\newline\urlprefix\url{http://jacow.org/ibic2016/papers/tupg72.pdf}

\bibitem{FE-I4:2012}
{FE-I4 Collaboration}, {The FE-I4B Integrated Circuit Guide}, 2nd Edition
  (2012).

\bibitem{Ahlburg:2016}
P.~Ahlburg, {Development of a FE-I4-based module for radiation monitoring with
  BEAST II during the commissioning phase of the Belle II detector}, Master's
  thesis, University of Bonn (2016).

\bibitem{Cuesta:2020bpl}
D.~Cuesta, J.~Baudot, G.~Claus, M.~Goffe, K.~Jaaskelainen, L.~Santelj,
  M.~Specht, M.~Szelezniak, I.~Ripp-Baudot, {Operation of a double-sided CMOS
  pixelated detector at a high intensity $e^+e^-$ particle collider}, Nucl.
  Instrum. Meth. A 967 (2020) 163862.
\newblock \href {http://arxiv.org/abs/2002.06941} {\path{arXiv:2002.06941}},
  \href {https://doi.org/10.1016/j.nima.2020.163862}
  {\path{doi:10.1016/j.nima.2020.163862}}.

\bibitem{Nomerotski:2011zz}
A.~Nomerotski, et~al., {PLUME collaboration: Ultra-light ladders for linear
  collider vertex detector}, Nucl. Instrum. Meth. A650 (2011) 208--212.
\newblock \href {https://doi.org/10.1016/j.nima.2010.12.083}
  {\path{doi:10.1016/j.nima.2010.12.083}}.

\bibitem{Baudot:2013pca}
J.~Baudot, et~al., {Optimization of CMOS pixel sensors for high performance
  vertexing and tracking}, Nucl. Instrum. Meth. A 732 (2013) 480--483.
\newblock \href {http://arxiv.org/abs/1305.0531} {\path{arXiv:1305.0531}},
  \href {https://doi.org/10.1016/j.nima.2013.06.101}
  {\path{doi:10.1016/j.nima.2013.06.101}}.

\bibitem{Jaegle:2019jpx}
I.~Jaegle, et~al., {Compact, directional neutron detectors capable of
  high-resolution nuclear recoil imaging} (2019).
\newblock \href {http://arxiv.org/abs/1901.06657} {\path{arXiv:1901.06657}},
  \href {https://doi.org/10.1016/j.nima.2019.06.037}
  {\path{doi:10.1016/j.nima.2019.06.037}}.

\bibitem{Thorpe:2021qce}
T.~N. Thorpe, S.~E. Vahsen, {Avalanche gain and its effect on energy resolution
  in GEM-based detectors} (6 2021).
\newblock \href {http://arxiv.org/abs/2106.15568} {\path{arXiv:2106.15568}}.

\bibitem{Schueler:2021}
J.~Schueler, S.~E. Vahsen, P.~M. Lewis, M.~T. Hedges, D.~Liventsev, F.~Meier,
  H.~Nakayama, A.~Natochii, T.~N. Thorpe, {Application of recoil-imaging time
  projection chambers to directional neutron background measurements in the
  SuperKEKB accelerator tunnel} (11 2021).
\newblock \href {http://arxiv.org/abs/2111.03841} {\path{arXiv:2111.03841}}.

\bibitem{Hedges:2021dgz}
M.~T. Hedges, S.~E. Vahsen, I.~Jaegle, P.~M. Lewis, H.~Nakayama, J.~Schueler,
  T.~N. Thorpe, {First 3D vector tracking of helium recoils for fast neutron
  measurements at SuperKEKB} (6 2021).
\newblock \href {http://arxiv.org/abs/2106.13079} {\path{arXiv:2106.13079}}.

\bibitem{Bassi:2021dno}
G.~Bassi, L.~Bosisio, P.~Cristaudo, M.~Dorigo, A.~Gabrielli, Y.~Jin,
  C.~La~Licata, L.~Lanceri, L.~Vitale, {Calibration of diamond detectors for
  dosimetry in beam-loss monitoring}, Nucl. Instrum. Meth. A 1004 (2021)
  165383.
\newblock \href {http://arxiv.org/abs/2102.03273} {\path{arXiv:2102.03273}},
  \href {https://doi.org/10.1016/j.nima.2021.165383}
  {\path{doi:10.1016/j.nima.2021.165383}}.

\bibitem{epics}
{Experimental Physics and Industrial Control System}, {Accessible at
  \url{https://epics.anl.gov/}}.

\bibitem{Pang:2019ses}
C.~G. Pang, P.~Bambade, S.~Di~Carlo, Y.~Funakoshi, D.~Jehanno, V.~Kubytskyi,
  M.~Masuzawa, Y.~Peinaud, C.~Rimbault, S.~Uehara, {A fast luminosity monitor
  based on diamond detectors for the SuperKEKB collider}, Nucl. Instrum. Meth.
  A931 (2019) 225--235.
\newblock \href {https://doi.org/10.1016/j.nima.2019.03.071}
  {\path{doi:10.1016/j.nima.2019.03.071}}.

\bibitem{css}
{Control System Studio}, {Accessible at \url{http://controlsystemstudio.org/}}.

\bibitem{ROOT}
R.~Brun, F.~Rademakers, {ROOT - An Object Oriented Data Analysis Framework}
  (1997).

\bibitem{SADHP}
{Strategic Accelerator Design}, available online at:
  \url{http://acc-physics.kek.jp/SAD/} (2021).

\bibitem{natochii2021improved}
A.~Natochii, S.~E. Vahsen, H.~Nakayama, T.~Ishibashi, S.~Terui, {Improved
  simulation of beam backgrounds and collimation at SuperKEKB} (2021).
\newblock \href {http://arxiv.org/abs/2104.02645} {\path{arXiv:2104.02645}}.

\bibitem{AGOSTINELLI2003250}
S.~Agostinelli, et~al.,
  \href{https://www.sciencedirect.com/science/article/pii/S0168900203013688}{Geant4—a
  simulation toolkit}, Nuclear Instruments and Methods in Physics Research
  Section A: Accelerators, Spectrometers, Detectors and Associated Equipment
  506~(3) (2003) 250--303.
\newblock \href {https://doi.org/https://doi.org/10.1016/S0168-9002(03)01368-8}
  {\path{doi:https://doi.org/10.1016/S0168-9002(03)01368-8}}.
\newline\urlprefix\url{https://www.sciencedirect.com/science/article/pii/S0168900203013688}

\bibitem{Kuhr_2018}
T.~Kuhr, C.~Pulvermacher, M.~Ritter, T.~Hauth, N.~Braun,
  \href{http://dx.doi.org/10.1007/s41781-018-0017-9}{{The Belle II Core
  Software}}, Computing and Software for Big Science 3~(1) (2018).
\newblock \href {https://doi.org/10.1007/s41781-018-0017-9}
  {\path{doi:10.1007/s41781-018-0017-9}}.
\newline\urlprefix\url{http://dx.doi.org/10.1007/s41781-018-0017-9}

\bibitem{the_belle_ii_collaboration_2021_5574116}
T.~B.~I. Collaboration, \href{https://doi.org/10.5281/zenodo.5574116}{{Belle II
  Analysis Software Framework (basf2): release-06-00-00}} (Oct. 2021).
\newblock \href {https://doi.org/10.5281/zenodo.5574116}
  {\path{doi:10.5281/zenodo.5574116}}.
\newline\urlprefix\url{https://doi.org/10.5281/zenodo.5574116}

\bibitem{rave}
W.~Waltenberger, IEEE Trans. Nucl. Sci. 58 (2011) 434.

\bibitem{SchreeckPhD:2020}
H.~Schreeck, {Commissioning and first data taking experience with the Belle II
  pixel vertex detector}, Ph.D. thesis, University of Göttingen, available
  online at: \url{http://hdl.handle.net/21.11130/00-1735-0000-0005-13F7-F}
  (2020).

\bibitem{RITTER201379}
A.~Ritter, et~al., {Investigations on radiation hardness of DEPFET sensors for
  the Belle II detector}, Nucl. Instrum. Methods Phys. Res., Sect. A 730 (2013)
  79 -- 83.
\newblock \href {https://doi.org/https://doi.org/10.1016/j.nima.2013.04.069}
  {\path{doi:https://doi.org/10.1016/j.nima.2013.04.069}}.

\bibitem{Schreeck2020}
H.~Schreeck, et~al., {Effects of gamma irradiation on DEPFET pixel sensors for
  the Belle II experiment}, Nucl. Instrum. Methods Phys. Res., Sect. A 959
  (2020) 163522.
\newblock \href {https://doi.org/10.1016/j.nima.2020.163522}
  {\path{doi:10.1016/j.nima.2020.163522}}.

\end{thebibliography}



\end{document}